\documentclass[12pt]{article}
\usepackage{amsmath}
\usepackage{graphicx}
\usepackage{enumerate}
\usepackage{natbib}
\usepackage{url} % not crucial - just used below for the URL
\usepackage{authblk}
\newcommand{\blind}{1}

\newcommand{\pperp}{{\ \perp\!\!\!\perp\ }}
\newcommand\setItemnumber[1]{\setcounter{enumi}{\numexpr#1-1\relax}}
\newtheorem{theorem}{Theorem}[section]
\newtheorem{thm}[theorem]{Theorem}
\usepackage {tikz}
\usetikzlibrary {positioning}
\definecolor {processblue}{cmyk}{0.96,0,0,0}

\begin{document}

\def\spacingset#1{\renewcommand{\baselinestretch}%
{#1}\small\normalsize} \spacingset{1}

%%%%%%%%%%%%%%%%%%%%%%%%%%%%%%%%%%%%%%%%%%%%%%%%%%%%%%%%%%%%%%%%%%%%%%%%%%%%%%

\if1\blind
{
  \title{\bf Bayesian model selection approach for colored graphical Gaussian models}
  \author{Qiong Li \textsuperscript{a} \ \ \
    Xin Gao  \thanks{ To whom correspondence should be addressed. Email: xingao@mathstat.yorku.ca}  \ \ \ H\'{e}l\`{e}ne Massam \thanks{\textsuperscript{*}Supported in part by the Natural Sciences and Engineering Research Council of Canada (NSERC) individual discovery grants.}\\
    \textsuperscript{a} School of Mathematics (Zhuhai), Sun Yat-sen University, Zhuhai, 519082, Guangdong, China\\
    \textsuperscript{*} \textsuperscript{\dag} Department of Mathematics and Statistics, York University, Toronto, M3J1P3, Canada}
  \maketitle
} \fi

\if0\blind
{
  \bigskip
  \bigskip
  \bigskip
  \begin{center}
    {\LARGE\bf Bayesian model selection approach for colored graphical Gaussian models}
\end{center}
  \medskip
} \fi

\bigskip
\begin{abstract}
We consider a class of colored graphical Gaussian models obtained by placing symmetry constraints on the precision matrix in a Bayesian framework. The prior distribution on the precision matrix is the colored $G$-Wishart prior which is the Diaconis-Ylvisaker conjugate prior.
In this paper, we develop a computationally efficient model search algorithm which combines linear regression with a
 double reversible jump Markov chain Monte Carlo (MCMC) method. The latter is to estimate the Bayes factors expressed as the ratio of posterior probabilities of two competing models. We also establish the asymptotic consistency property of the  model selection procedure based on the Bayes factors. Our procedure avoids an exhaustive search which is computationally impossible.
  Our method is illustrated with simulations and a real-world application with a protein signalling data set.
\end{abstract}

\noindent%
{\it Keywords:}  Auxiliary variable MCMC algorithm; Colored $G$-Wishart distribution; Double reversible jump
\vfill

\newpage
\spacingset{1.5} % DON'T change the spacing!
\section{Introduction}
\label{sec:intro}
Graphical models provide an effective way to determine conditional independence relationships in high-dimensional data. Application of these models arise in the study of gene expression data \citep{Dob04}, natural language processing \citep{Jung96} and image analysis \citep{Li01}. A graphical model is used to display the most significant interactions between random variables. It is thus a quite viable modelling tool for a large variety of real-life datasets. A pertinent example we focus on in this paper is a cell signalling data set collected by \citet{Sachs05}, which consists of 7,466 measurements on 11 phosphorylated proteins involved in primary human immune system cells. Measurements were performed using flow cytometry, which is a popular cell biology technique that produces large samples for measurements of the total amount of proteins. The Bayesian method for colored graphical models selection we develop in this paper allows us to capture the complex pattern of conditional associations and symmetric structures that exist among the proteins.

\citet{Hoj08} introduce so-called colored graphical models by adding equality constraints on the entries of  the correlation matrix $\mathbf{R}$, or on the entries of the precision matrix $\textbf{K}$. In this paper, we will work on the class of  colored graphical models obtained by imposing  arbitrary equality constraints on $\textbf{K}$. Such a model can be represented by a colored graph, where edges or vertices have the same coloring if the corresponding entries of $\textbf{K}$ are restricted to being identical. Using a colored graphical model typically allows for a reduction in the dimension of the parameter space. Thereby these models can be applied to problems where the number of variables $p$ is substantially larger than the number of observations $n$.

We now briefly review the various computational methods to perform model selection in the class of  uncolored graphical Gaussian models. \citet{Dob11} developed a reversible jump algorithm \citep{Green95}. This approach assumes that we can compute the ratio of prior normalizing constants which, in small to moderate dimensions, can be estimated using the Monte Carlo method of \citet{Ata05}.  But, as noted by \citet{Wang12}, this approximation   cannot be used  in high dimensions. \citet{Liang10} derived a double Metropolis-Hastings sampler, which is an extension of the exchange algorithm \citep{Murray06} for simulating from distributions with intractable normalizing constants. This auxiliary variable algorithm avoids the calculation of the normalizing constants altogether and removes the need for exact sampling. \citet{Wang12} adopted the idea from the double Metropolis-Hasting algorithm to perform graphical Gaussian model selection. \citet{Len13} proposed a direct sampler for the $G$-Wishart distribution and developed a new transdimensional double reversible jump algorithm which integrates the exchange algorithm \citep{Murray06} with the reversible jump MCMC \citep{Green95}.

To the best of our knowledge, there is no efficient Bayesian method for model selection in the class of colored graphical Gaussian models. When searching this class of models, one is faced with two main problems.
One is the efficient computation or estimation of normalizing constants of the colored $G$-Wishart distribution. The other is the super-exponential growth of the dimension of the colored graph space in $p$. For example, when $p$ is 4 or 5, the number of uncolored graphs in the space is 64 and 1,024 respectively, whereas the number of colored graphs is 13,155 or 35,285,640 \citep[see][]{Gehrmann11}.

In this paper, we construct a novel model search procedure which couples linear regression with the double reversible jump scheme. We use linear regression to add significant edges. The model $G^*$ with an additional edge is then compared to the current model $G$ using the Bayes factor $p(G^*|\textbf{X})/p(G|\textbf{X})$ which itself is computed
 with the help of the double reversible jump MCMC algorithm. The double reversible jump consists of two moves: one move is under the posterior distribution on the state space $(G, K)$, and the other is an auxiliary move under the prior distribution on the same state space. The double reversible jump algorithm allows for the cancellation of the prior normalizing constants in the expression of the acceptance probability of the chain on $(G, K)$. We thus avoid computing these quantities which are the usual computational stumbling blocks in graphical Gaussian model selection.
%The double reversible jump approach introduces an auxiliary variable $K'$ and constructs a Markov chain whose stationary distribution is taken to be the joint posterior distribution of $(G,K)$ and $(G',K')$. The Bayes factor is approximated by the ratio of two marginal posterior distributions of $G$ and $G'$ given the data. The proposed procedure avoids the global moves of Markov chains in the colored graph space and makes the procedure simpler but highly efficient. This approach adds significant edges through linear regression, then compares the current graph and the candidate graph using the double reversible jump MCMC algorithm. Our model selection approach avoids the calculation of normalizing constants of the colored $G$-Wishart distribution and saves the computing time for the setting of large $p$.

The rest of the paper is organized as follows. Section 2 formally introduces colored graphical models and the colored $G$-Wishart distribution. Section 3 discusses the model search  approach based on linear regression. Section 4 describes our reversible jump MCMC and double reversible jump MCMC algorithms.
%Section 5 computes normalizing constants for some specific colored graphs in order to evaluate the efficiency of our method.
Section 5 concentrates on the property of model consistency for the proposed model selection algorithm. Section 6 demonstrates the performance of our method through several numerical experiments. Finally, Section 7 presents a real-world application.

\section{Preliminaries and notation}

%\subsection{Colored $G$-Wishart distribution}
We will start by reviewing some of the basic concepts related to colored graphical Gaussian models. For a detailed description, the reader is referred to \citet{Hoj08} and \citet{Massam18}. We consider an undirected graph $G = (V,E)$ where $V$ is the set of vertices and $E\subset V\times V$ is the set of undirected edges. Let $X = (X_{v}, v \in V)$ be a random vector following a $p$-dimensional multivariate normal distribution $N_{p}(0,\textbf{K}^{-1})$.

A colored graphical Gaussian model with respect to a colored graph $G$ is constructed by setting some of the off-diagonal elements of $\textbf{K}$ to be zero, and some of the diagonal elements or off-diagonal elements of $\textbf{K}$ to be identical.
We say that $\textbf{K}$  belongs to the cone $P_{G}$ of symmetric positive definite matrices with  zero or equality constraints on the entries of $\textbf{K}$.
If $K_{v_{1},v_{2}}=0$ for $(v_{1},v_{2}) \notin E$, then the corresponding variables $X_{v_{1}}$ and $X_{v_{2}}$ are conditionally independent giving the remaining variables, which denoted by
\begin{eqnarray*}
X_{v_{1}}\pperp X_{v_{2}} \mid X_{V\setminus\{v_{1},v_{2}\}}.
\end{eqnarray*}
This is called the pairwise Markov property relative to the graph $G$. If the diagonal elements of $\textbf{K}$ or the off-diagonal elements are identical, then the corresponding vertices or edges, respectively,  are in the same color class. The equality on $\textbf{K}$ imposes the symmetric structure of the underlying graph $G$.

The prior distribution on $\textbf{K}$ is the Diaconis-Ylvisaker conjugate prior for this model and is called as the colored $G$-Wishart distribution \citep{Massam18} defined on $P_{G}$ and with the density
\begin{eqnarray}\label{prior}
p(\textbf{K}|\delta, \textbf{D}, G) = \frac{1}{I_{G}(\delta, \textbf{D})}|\textbf{K}|^{(\delta-2)/2}\exp\big\{-\frac{1}{2} <\textbf{K},\textbf{D}>\big\}{\bf 1}_{ P_{G}}(\textbf{K}),
\end{eqnarray}
where $\delta > 0$, $\textbf{D}$ is a symmetric positive definite matrix, $<\mathbf{A}, \mathbf{B}>$ is the trace inner product, ${\bf 1}_{ P_{G}}$ is the indicator function of $P_G$
and $I_{G}(\delta, \textbf{D})$ is the normalizing constant, namely,
$$I_{G}(\delta, \textbf{D})=\int |\textbf{K}|^{(\delta-2)/2}\exp\big\{-\frac{1}{2} <\textbf{K}, \textbf{D}>\big\}{\bf 1}_{P_{G}}(\textbf{K})d\textbf{K}.$$
The goal of the proposed reversible jump MCMC and  double reversible jump MCMC is to generate samples from the joint posterior density $p(G,\textbf{K}|\textbf{X})$ given the data $\textbf{X}=(x^{1},\ldots, x^{n})$.
The posterior density is
\begin{eqnarray*}
p(G, \textbf{K}|\textbf{X})& = & p(\textbf{X}|G,\textbf{K})p(\textbf{K}|G)p(G)\\
&\propto &\frac{1}{I_{G}(\delta, \textbf{D})}|\textbf{K}|^{(n+\delta-2)/2}\exp\big\{-\frac{1}{2} <\textbf{K},\textbf{S}+\textbf{D}>
\big\}p(G){\bf 1}_{P_{G}}(\textbf{K}),
\end{eqnarray*}
where  $\textbf{S}=\sum\limits^{n}_{i=1}(x^{i})(x^{i})^\top,$ and $p(G)$ denotes the prior distribution on the class of graphs considered.

Let $\textbf{K}=\mathbf{\Phi}^\top\mathbf{\Phi}$, where $\mathbf{\Phi}=(\Phi_{ij})_{1\leq i\leq j\leq p}$ is an upper triangular matrix with real positive diagonal elements, be the Cholesky decomposition of the matrix $\textbf{K}$. Let $\mathcal{V}=\{V_1,\dots,V_t\}$ and $\mathcal{E}=\{E_1,\dots,E_k\}$ denote the set of vertex color classes and edge color classes, respectively. Denote
$$v_{u}(G)=\min\{(i,j)\in u | i\leq j, \; u \in \mathcal{V}\cup \mathcal{E} \},$$
where the minimum is defined according to the lexicographical order and define
$$v(G)=\bigcup \limits_{u \in \mathcal{V}\cup \mathcal{E}}v_{u}(G).$$ The vertices and edges $(i,j) \in v(G)$ are called free vertices and free edges, respectively. The remaining vertices and edges are called non-free.
\citet{Massam18} proved that
the Jacobian of the change of variables $\textbf{K}^{v(G)}\rightarrow \mathbf{\Phi} ^{v(G)}$ is
$$|J(\textbf{K}^{v(G)}\rightarrow \mathbf{\Phi}^{v(G)})|=2^{|V_{G}|}\prod\limits^{p}_{i=1}\Phi_{ii}^{p-i+1-v_i^G}$$
where $|V_{G}|$ is the number of vertex color classes of $G$ and $v_i^G$ is the number of $j\in \{i,\ldots,p\}$ such that $(i,j)\not\in v(G)$. Then the posterior distribution of  $(G, \mathbf{\Phi})$ can be rewritten as
\begin{eqnarray}\label{posterior}
p(G,\mathbf{\Phi}|\textbf{X}) &=& p(\textbf{X}|G,\textbf{K})p(\textbf{K}|G)|J(\textbf{K}^{v(G)}\rightarrow \mathbf{\Phi}^{v(G)})|p(G)\nonumber\\
&\propto& \frac{2^{|V_{G}|}}{I_{G}(\delta,\textbf{D})}\prod\limits_{i=1}^{p}\Phi_{ii}^{n+\delta+p-i-1-v_{i}^{G}}\exp\{-\frac{1}{2}<\mathbf{\Phi}^\top\mathbf{\Phi},\textbf{D}+\textbf{S}>\}p(G).
\end{eqnarray}
We write $\textbf{K}^{v(G)}=\{K_{ij}:(i,j)\in v(G)\}$ for the free elements of $\textbf{K}$ and $\mathbf{\mathbf{\Phi}}^{v(G)}=\{\Phi_{ij}:(i,j)\in v(G)\}$  for the corresponding free elements of  ${\mathbf{\Phi}}$.

\section{Algorithm for model selection}

As mentioned in the introduction, the number of possible colored graphical models grows exponentially with the number $p$ of variables for a random variable $\textbf{X}=(X_v, v=1,\ldots,p)$. It is therefore computationally impossible to travel through the space of graphs and compare all models in the space. We thus propose here a new search method for model selection in the class of colored Gaussian graphical models. This method combines linear regression to move from one model to another, with the double reversible jump algorithm to compute the ratio of  posterior probabilities of the two models that we compare.
 Recall that if we partition $\textbf{X}$ into $X_j$ and $X_{-j}=(X_1,\ldots, X_{j-1}, X_{j+1},\ldots, X_p)$, then the conditional distribution of $X_j$ given $X_{-j}$ is Gaussian and the mean $E(X_j\mid X_{-j})$  is a linear combination of the components of $X_{-j}$ \citep{Hastie09,Meinshausen06}.

 The algorithm starts with the regression step and then attempts to classify the new edges into color classes. It goes as follows.

%The search procedure moves to vertex constraints addition step when there are no edges to be added. The algorithm terminates and returns the best colored graph when there are no edges and colored constraints to be added.
\begin{itemize}
  \item[{\bf Step 1}] Start  with the graph $G^{[1]}$ with $p$ vertices and no edges. All vertices are free in $G^{[1]}$.

  \item[{\bf  Step 2}] Let $G^{[t]}$ be the current graph. Repeat for $j=1,2,\cdots,p.$
\begin{itemize}
  \item[ 2.a] Run the following linear regression to search for potential edges to be added for the graph $G^{[t]}$.
\begin{equation}\label{regression}
X_j\sim \beta^j_1X_1+\cdots+\beta^j_{j-1}X_{j-1}+\beta^j_{j+1}X_{j+1}+\cdots+\beta^j_p X_p.
\end{equation}
Set $\beta^j_{G^{[t]}} = \{\beta^j_i|(i,j) \in G^{[t]}\}$ and find all the significant coefficients $\beta^j_i$ with $p$-values less than $\alpha$ in linear regression  among  $\{\beta^j_1,\cdots,\beta^j_{j-1},\beta^j_{j+1},\beta^j_p\}\setminus \beta^j_{G^{[t]}}.$

\item[2.b] Order the edges from the most significant to the least significant. For each of the edge, we decide whether to accept that edge or not and also determine its color on the basis of the Bayes factor  as follows.
  \begin{itemize}

\item[2.b.i] Let $G^{*}$ be the graph obtained by adding the edge $(i,j)$  to  $G^{[t]}$. Implement model selection between the candidate $G^{*}$ and the current graph $G^{[t]}$ by estimating the  Bayes factor $p(G^*|\textbf{X})/p(G^{[t]}|\textbf{X})$ with the help of the double reversible jump algorithm. If the ratio is greater than one, we accept the graph $G^*$ and set $G^{[t+1]}=G^{*}$. Otherwise, we accept the graph $G^{[t+1]}=G^{[t]}$.
  \item[ 2.b.ii]   We try to merge the new added edge into the existing color classes. Order the edges in
  $$v(G^{[t+1]})\cap {\mathcal E}=\{u_1,\ldots, u_k, k=|v(G^{[t+1]})\cap {\mathcal E}|\}$$
  in the lexicographic order.
  For $l=1,\ldots,k,$
  \begin{enumerate}
  \item[($\alpha$)]
   Let $G^{**}$ be the graph obtained from $G^{[t+1]}$ by setting the the new edge  $(i,j)$  accepted (if so) in Step 2.b.i to be in the same class color as $u_l$.
Implement model selection between $G^{**}$ and the current graph $G^{[t+1]}$ by estimating the Bayes factor $p(G^{**}|\textbf{X})/p(G^{[t+1]}|\textbf{X})$.
  If $G^{**}$ is returned, set $G^{[t+2]}=G^{**}$, exit the current color determination step and go back to Step 2.b with the next most significant edge found in Step 2.a.
  \item[($\beta$)]
  If $(i,j)$ is not accepted in $u_l$ and $l\not =k$, let $l=l+1$ and go back to Step 2.b.ii($\alpha$).
  \item[($\gamma$)] If $l=k$ and $(i,j)$ cannot be merged into any  of the existing color classes, then $(i,j)$  becomes an additional member of $v(G^{[t+1]})\cap {\mathcal E}$ and represents a new color class of which it is the only element and the graph remains $G^{[t+2]}=G^{[t+1]}$.  Go back to Step 2.b with the next most significant edge found in Step 2.a.
 \end{enumerate}
\end{itemize}
\end{itemize}
\end{itemize}

\begin{enumerate}
 \item[{\bf  Step 3}] We determine the color classes for all vertices. Start with all vertices free. Repeat for $k=2,\cdots,p$.

\item[ 3.a]  Order the vertices in
  $$v(G^{[t+2]})\cap {\mathcal V}=\{u_1,\ldots, u_m, m=|v(G^{[t+2]})\cap {\mathcal V}|\}$$
  in the lexicographic order.
 For $l=1,\ldots,m,$
 \begin{enumerate}
\item[($\alpha$)]
Let $G^{**}$ be the graph obtained from $G^{[t+2]}$ by setting the vertex  $k$   to be in the same class color as $u_l$.
Compute the Bayes factor $p(G^{**}|\textbf{X})/p(G^{[t+2]}|\textbf{X})$ with the help of the double reversible jump algorithm.
 If the Bayes factor is larger than  one, the proposed vertex color merging is accepted, we set $G^{[t+3]}=G^{**}$. We then exit the current color determination step, set $k=k+1$ and go back to Step 3.
 \item[($\beta$)]If the Bayes factor is less than  one, we proceed to the next existing vertex color class and let $l=l+1$.
  \item[($\gamma$)]
 If $l=m$ and  vertex $k$ cannot be merged into any existing vertex color class,  then $k=(i,i)$  becomes an additional member of $v(G^{[t+2]})\cap {\mathcal V}$ and represents a new color class of which it is the only element and the graph remains $G^{[t+3]}=G^{[t+2]}$.

\end{enumerate}
\end{enumerate}

The search algorithm above starts with an uncolored graph without any edge and then incrementally adds significant edges by linear regression. Each successful edge addition step is followed by a color determination step for the new edge. The acceptance of a new model is always decided based on the value of the Bayes factor for the proposed model versus the current model. We illustrate the algorithm on a toy example. Suppose that we have a current graph with two edge color classes in graph $G$, denoted as red and  green in that order. Through linear regression, we add a new edge to the graph. If this new graph is accepted, the new edge is automatically assigned a new color black. Then we try to merge the black edge into the red class first. If accepted, the black edge is changed to a red edge and we consider the next new most significant edge. The current graph has two edge color classes, red and green. If rejected, we then try to merge the black edge into the green class. If accepted, the black edge is changed to a green edge and we consider the next new most significant edge. The current graph still has two edge color classes, red and green. If rejected, the current graph keeps three edge color classes.  After considering the edge color classes, we focus on the vertex color classes. We try to merge free vertices into existing color classes. We accept or reject the proposed merge based on the Bayes factor of the proposed new model and the current model.
Our algorithm  avoids searching through the whole colored graph space, which, as mentioned above, is computationally impossible even for very moderate dimensions.

\section{Estimation of Bayes factors}

To evaluate Bayes factors such as $p(G^*|\textbf{X})/p(G^{[t]}|\textbf{X}),$ we aim to simulate a MCMC
chain on the posterior distribution of $P(G,\mathbf{\Phi}|\textbf{X}),$ where $\textbf{K}=(\mathbf{\Phi})^\top\mathbf{\Phi}$  follows the colored $G$-Wishart distribution $CW_G(\delta+n,\textbf{D}+\textbf{S}).$ We sample $G$ from the two neighboring
colored graphs $\{G^*, G^{[t]}\}$ and each graph is assumed to have equal prior. Based on this sampling of $G$ from the simulated posterior distribution, we count the proportion of samples belonging to graph $G^*$ and $G^{[t]}$ respectively as the empirical estimates of their marginal posterior distributions $P(G^*|\textbf{X})$ and $P(G^{[t]}|\textbf{X})$. Then we estimate the Bayes factor based on these proportions. We extend the ideas from the reversible jump MCMC \citep{Dob11}
and the double reversible jump \citep{Len13} for uncolored graphical models to the colored graphical models and colored $G$-Wishart distribution.

Given a colored graph $G$, a neighbor of $G$ is a graph obtained by deleting or adding one edge in $G,$ or a graph obtained by deleting or adding one vertex color
class or one edge color class. Here deleting a color class means that an entry $(i,j)$
that was previously free becomes a non-free one, that is, it joins an existing color class
$u\in{\cal V}\cup {\cal E}$. Adding a color class means that an entry $(i, j)$ that belonged to a color
class $u\in{\cal V}\cup {\cal E}$ becomes free, creating a color class of its own and thus increasing the
cardinality of ${\cal V}\cup {\cal E}$. The chain moves between two neighboring colored graphs which
are in two different dimensions of the parameter space.

\subsection{Reversible jump MCMC}

We describe a reversible jump sampler based on a colored $G$-Wishart prior for $\textbf{K}$. This approach requires the calculation of the normalizing constants of the $G$-Wishart priors corresponding to the current and the candidate graphs. It can thus be used only for some small specific graphs for which we can compute the normalizing constants.
%The element of $K$  in the two colored graphs becomes free, or constrained to zero or to be equal by adding or deleting an edge, or changing the color of edges or vertices.
Let us, now, present the details of the reversible jump algorithm for different cases. We denote the current state of the chain by $(\textbf{K}^{[t]},G^{[t]})$ and the next state by $(\textbf{K}^{[t+1]},G^{[t+1]})$. For a given colored graph $G^{[t]}$, we make use of the Monte Carlo method of \citet{Massam18} to obtain the samples $\textbf{K}^{[t]}$ from the colored $G$-Wishart distribution. Let $G^*$ be the candidate colored graph. Let $\textbf{K}^{[t]}=(\mathbf{\Phi}^{[t]})^\top\mathbf{\Phi}^{[t]}$ and $\textbf{K}^{*}=(\mathbf{\Phi}^{*})^\top\mathbf{\Phi}^{*}$ be the respective Cholesky decomposition of the precision matrices.  We update one entry of $\mathbf{\Phi}$ at a time. All the non-updated free elements of $\mathbf{\Phi}^{[t]}$ coincide with the corresponding ones of $\mathbf{\Phi}^*$.  The non-free elements of $\mathbf{\Phi}^{*}$ are determined through the completion operation by Proposition 1 in \citet{Massam18}.

Case 1. The candidate $G^*$ is obtained by  changing the non-free edge $(i,j)$ of $G^{[t]}$ to a free one. We consider two different scenarios. In the first scenario, $G^*$ is obtained by adding the edge $(i,j)$ on $G^{[t]}$. In the second scenario, $G^*$ is obtained by removing a color constraint on the edge $(i,j)$.
For both scenarios, we sample
$\gamma \sim N(\Phi_{ij}^{[t]},\sigma)$ which is a proposal distribution for the updated $\Phi^{*}_{ij}$ and set $\Phi^{*}_{ij} = \gamma$. The Markov chain moves to $(G^*, \textbf{K}^*)$ with probability
$\min\{R_{0}^+,1\}$ where $R_{0}^+$ is
\begin{eqnarray*}
&&\frac{p(G^*,\mathbf{\Phi}^*|\textbf{X})}{p(G^{[t]},\mathbf{\Phi}^{[t]}|\textbf{X})}*\frac{1}{\frac{1}{\sqrt{2\pi}\sigma}\exp\{-\frac{1}{2\sigma^2}(\Phi^*_{ij}-\Phi^{[t]}_{ij})^2\}}\\&=&
\frac{ \frac{2^{|V_{G^*}|}}{I_{G^*}(\delta,\textbf{D})}\prod\limits_{i=1}^{p}(\Phi^{*}_{ii})^{n+\delta+p-i-1-v_{i}^{G^*}}\exp\{-\frac{1}{2}<(\mathbf{\Phi}^*)^\top\mathbf{\Phi}^*,\textbf{D}+\textbf{S}>\}p(G^*)}{ \frac{2^{|V_{G^{[t]}}|}}{I_{G^{[t]}}(\delta,\textbf{D})}\prod\limits_{i=1}^{p}(\Phi^{[t]}_{ii})^{n+\delta+p-i-1-v_{i}^{G^{[t]}}}\exp\{-\frac{1}{2}<(\mathbf{\Phi}^{[t]})^\top\mathbf{\Phi}^{[t]},
\textbf{D}+\textbf{S}>\}p(G^{[t]})}\\
&&*\frac{\sqrt{2\pi}\sigma}{\exp\{-\frac{1}{2\sigma^2}(\Phi^*_{ij}-\Phi^{[t]}_{ij})^2\}}.
\end{eqnarray*}

Case 2. The candidate $G^*$ is obtained by changing the non-free vertex $(i,i)$ on $G^{[t]}$ to a free one. We sample
$\gamma$ from a proposal distribution $N^{+}(\Phi_{ii}^{[t]}; \sigma, 0, \infty)$ which is the normal distribution truncated below at zero, with mean $\Phi_{ii}^{[t]}$ and set $\Phi^{*}_{ii} = \gamma$. The Markov chain moves to $(G^*, \textbf{K}^*)$ with probability
$\min\{R_{v}^+,1\}$ where $R_{v}^+$ is
\small\begin{eqnarray*}
\frac{\frac{2^{|V_{G^*}|}}{I_{G^*}(\delta,\textbf{D})}\prod\limits_{i=1}^{p}(\Phi^{*}_{ii})^{n+\delta+p-i-1-v_{i}^{G^*}}\exp\{-\frac{1}{2}<(\mathbf{\Phi}^*)^\top\mathbf{\Phi}^*,\textbf{D}+\textbf{S}>\}p(G^*)}{ \frac{2^{|V_{G^{[t]}}|}}{I_{G^{[t]}}(\delta,\textbf{D})}\prod\limits_{i=1}^{p}(\Phi^{[t]}_{ii})^{n+\delta+p-i-1-v_{i}^{G^{[t]}}}\exp\{-\frac{1}{2}<(\mathbf{\Phi}^{[t]})^\top\mathbf{\Phi}^{[t]},
\textbf{D}+\textbf{S}>\}p(G^{[t]})}*\frac{1}{f(\Phi^*_{ii};\Phi^{[t]}_{ii}, \sigma,0,+\infty)}
\end{eqnarray*}
where $f(*;\mu,\sigma,0,+\infty)$ is the density function for the truncated normal distribution.

\subsection{Double reversible jump MCMC}

The reversible jump MCMC in subsection 4.1 assumes that we can compute the normalizing constants $I_{G}(\delta, \textbf{D})$. In Section 6, we compute the normalizing constant of some specific colored graphs. But, in general, we do not know how to obtain the analytic expression of these constants or even how to estimate them efficiently. To
circumvent this problem, we will use the double reversible jump MCMC algorithm
which avoids the calculation of the normalizing constants. %Therefore, to compute the  \textcolor{blue}{ Bayes factor}, we will use the double reversible jump MCMC algorithm which we are now going to describe.
%Liang (2013) proposed the double MH sampler for distributions with intractable normalizing constants. Lenkoski (2013) derived a sample scheme for uncolored $G$-Wishart distribution combing the double MH method and reversible jump MCMC. We here borrow this ideas to estimate Bayes factors for colored models and call this new approach double reversible jump method.

We sample from the posterior distribution $p(G,\mathbf{\Phi}|\textbf{X})$, where $\textbf{K}=(\mathbf{\Phi})^\top\mathbf{\Phi}$  follows the colored $G$-Wishart distribution $CW_G(\delta+n,\textbf{D}+\textbf{S}).$ Then we introduce the auxiliary variables $\tilde{G}$ and $\mathbf{\Omega},$ which share the same state spaces as $G$ and $\mathbf{\Phi}$, and the positive definite matrix $\mathbf{\Psi}=(\mathbf{\Omega})^\top\mathbf{\Omega}$  which follows the colored $G$-Wishart distribution $CW_{\tilde{G}}(\delta,\textbf{D}).$

We can simulate $(G,\mathbf{\Phi},\tilde{G},\mathbf{\Omega})$ as follows:

Case 1: If $G^*$ is obtained by adding one edge $(i,j)$ to the current graph $G^{[t]}$. Let $G_{1} = G^{[t]}$ and $G_{2} = G^{*}$.
\begin{enumerate}
\item[1.1] Sample $\textbf{K}^{[t]} \sim CW_{G_{1}}(\delta+n, \textbf{D}+\textbf{S})$ with $\textbf{K}^{[t]} = (\mathbf{\Phi}^{[t]})^\top \mathbf{\Phi}^{[t]}$ using the Monte Carlo method of \citet{Massam18}.
\item[1.2] Sample $\Phi^*_{ij} \sim N(\Phi^{[t]}_{ij}, \sigma),$ which is the proposal distribution of the updated $\Phi^*_{ij}$ with an arbitrary $\sigma.$  Let all free elements in $\mathbf{\Phi}^*$ take the same values as in $\Phi^{[t]}$ except for $\Phi_{ij}^*.$
\item[1.3] Sample $\mathbf{\Psi}^{[t]}=(\mathbf{\Omega}^{[t]})^\top\mathbf{\Omega}^{[t]}$ from $CW_{G_{2}}(\delta, \textbf{D})$ using the Monte Carlo method of \citet{Massam18}.
\item[1.4] Let the free elements in $\mathbf{\Omega}^{*}$
 take the same values as in $\mathbf{\Omega}^{[t]}$ except for $\Omega^{*}_{ij}$ and
\[ \Omega^{*}_{ij} =
  \begin{cases}
    0       & \quad \text{if} \quad i=1 \\
    -\frac{\sum\limits^{i-1}_{k=1}\Omega^{[t]}_{ki}\Omega^{[t]}_{kj}}{\Omega^{[t]}_{ii}}  & \quad \text{otherwise.}\\
  \end{cases}
\]

\item[1.5] Accept the move from $(G^{[t]},\mathbf{\Phi}^{[t]},\tilde{G}^{[t]}, \mathbf{\Omega}^{[t]})$ to $(G^*,\mathbf{\Phi}^*,\tilde{G}^*, \mathbf{\Omega}^*)$ with probability $\min\{1, r^{+}_0\}$ where
\begin{eqnarray*}
r^{+}_0&= & \frac{\frac{\prod\limits_{i=1}^{p}(\Phi^{*}_{ii})^{n+\delta+p-i-1-v_{i}^{G_{2}}}\exp\{-\frac{1}{2}<(\mathbf{\Phi}^*)^\top\mathbf{\Phi}^*,\textbf{D}+\textbf{S}>\}}{ \prod\limits_{i=1}^{p}(\Phi^{[t]}_{ii})^{n+\delta+p-i-1-v_{i}^{G_{1}}}\exp\{-\frac{1}{2}<(\mathbf{\Phi}^{[t]})^\top\mathbf{\Phi}^{[t]},\textbf{D}+\textbf{S}>\}}*\frac{1}{\exp\{-\frac{1}{2\sigma^2}(\Phi^*_{ij}-\Phi^{[t]}_{ij})^2\}}}
{\frac{\prod\limits_{i=1}^{p}(\Omega^{[t]}_{ii})^{\delta+p-i-1-v_{i}^{G_{2}}}\exp\{-\frac{1}{2}<(\mathbf{\Omega}^{[t]})^\top\mathbf{\Omega}^{[t]},\textbf{D}>\}}{ \prod\limits_{i=1}^{p}(\Omega^{*}_{ii})^{\delta+p-i-1-v_{i}^{G_{1}}}\exp\{-\frac{1}{2}<(\mathbf{\Omega}^{*})^\top\mathbf{\Omega}^{*},\textbf{D}>\}}*\frac{1}{\exp\{-\frac{1}{2\sigma^2}(\Omega^{[t]}_{ij}-\Omega^{*}_{ij})^2\}}}. \\
\end{eqnarray*}
\end{enumerate}
The term in the numerator is for  moving  from $(G_{1},\mathbf{\Phi}^{[t]})$ to $(G_{2},\mathbf{\Phi}^{*})$ where $\mathbf{\Phi}^{[t]}$ and $\mathbf{\Phi}^{*}$ follow the posterior distribution. The term in the denominator is for moving  from $(G_{2},\mathbf{\Omega}^{[t]})$ to $(G_{1},\mathbf{\Omega}^{*})$ where $\mathbf{\Omega}^{[t]}$ and $\mathbf{\Omega}^{*}$ follow the prior distribution.

Case 2: If $G^*$ is obtained by changing one non-free edge $(i,j)$ to a free edge from $G^{[t]}$ or, equivalently by adding one edge color class in $G^{[t]}$. Steps 2.1--2.3 and 2.5 are the same with those in Case 1.
\begin{enumerate}
\item[2.4.] Let the free elements in $\mathbf{\Omega}^{*}$ take the same values as in $\mathbf{\Omega}^{[t]}$ except for $\Omega^{*}_{ij}$ and
\[ \Omega^{*}_{ij} =
  \frac{\Omega^{[t]}_{i_{u}j_{u}}\Omega^{[t]}_{i_{u}i_{u}}+\sum\limits^{i_{u}-1}_{k=1}\Omega^{[t]}_{ki_{u}}\Omega^{[t]}_{kj_{u}}-
\sum\limits^{i-1}_{k=1}\Omega^{[t]}_{ki}\Omega^{[t]}_{kj}}{\Omega^{[t]}_{ii}},
\]
\end{enumerate}
where $u \in \mathcal{V} \cup \mathcal{E}$, $(i_{u},j_{u})=\min\{(i,j) \in u: i\leq j\}$ in the lexicographical order.

Case 3: If $G^*$ is obtained by changing one non-free vertex $(i,i)$ to a free vertex from $G^{[t]}$ or, equivalently adding one vertex color class in $G^{[t]}$. Steps 3.1 and 3.3 are the same as Steps 1.1 and 1.3.
\begin{enumerate}
\setItemnumber{2}
\item[3.2] Sample $\Phi^*_{ii}$ from the proposal distribution  $f(\Phi^*_{ii};\Phi^{[t]}_{ii}, \sigma,0,+\infty)$, the normal distribution truncated below at zero, and let the free $\mathbf{\Phi}^*=\mathbf{\Phi}^{[t]}$ except for $\Phi^*_{ii}$.
\setItemnumber{4}
\item[3.4] Let the free elements in $\mathbf{\Omega}^{*}$ take the same values as in $\mathbf{\Omega}^{[t]}$ except for $\Omega^{*}_{ii}$ and
\[ \Omega^{*}_{ii} =
  |(\Omega^{[t]}_{i_{u}i_{u}})^2+\sum\limits^{i_{u}-1}_{k=1}(\Omega^{[t]}_{ki_{u}})^2-\sum\limits^{i-1}_{k=1}(\Omega^{[t]}_{ki})^2|^{\frac{1}{2}}.
\]
\setItemnumber{5}
\item[3.5] Accept the move from $(G^{[t]},\mathbf{\Phi}^{[t]},\tilde{G}^{[t]}, \mathbf{\Omega}^{[t]})$ to $(G^*,\mathbf{\Phi}^*,\tilde{G}^*, \mathbf{\Omega}^*)$ with probability $\min\{1, r^{+}_v\}$ where
\begin{eqnarray*}
r^{+}_v &=& \frac{\frac{\prod\limits_{i=1}^{p}(\Phi^{*}_{ii})^{n+\delta+p-i-1-v_{i}^{G_{2}}}\exp\{-\frac{1}{2}<(\mathbf{\Phi}^*)^\top\mathbf{\Phi}^*,\textbf{D}+\textbf{S}>\}}{ \prod\limits_{i=1}^{p}(\Phi^{[t]}_{ii})^{n+\delta+p-i-1-v_{i}^{G_{1}}}\exp\{-\frac{1}{2}<(\mathbf{\Phi}^{[t]})^\top\mathbf{\Phi}^{[t]},\textbf{D}+\textbf{S}>\}}*\frac{1}{f(\Phi^*_{ii};\Phi^{[t]}_{ii}, \sigma,0,+\infty)}}{\frac{\prod\limits_{i=1}^{p}(\Omega^{[t]}_{ii})^{\delta+p-i-1-v_{i}^{G_{2}}}\exp\{-\frac{1}{2}<(\mathbf{\Omega}^{[t]})^\top\mathbf{\Omega}^{[t]},\textbf{D}>\}}{ \prod\limits_{i=1}^{p}(\Omega^{*}_{ii})^{\delta+p-i-1-v_{i}^{G_{1}}}\exp\{-\frac{1}{2}<(\mathbf{\Omega}^{*})^\top\mathbf{\Omega}^{*},\textbf{D}>\}}*\frac{1}{f(\Omega^{[t]}_{ii};\Omega^{*}_{ii},\sigma,0,+\infty)}}. \\
\end{eqnarray*}
\end{enumerate}
\section{Model selection consistency}

Denote $G_T$ as the true graph that generates the data $\textbf{X},$ and $G_s$ as any competitor graph that can be used to model $\textbf{X}.$
 In this section, we are going to show that, as the sample size increases, the probability that the Bayes factor selects the true graph $G_T$ over the competitor graph $G_s$ converges to one.

The competitor graphs fall into two categories: an underfitting graph, denoted by $G_{-}$ with missing edges or incorrect partitions of the edges into different color classes and
an overfitting graph denoted by $G_{+}$ with more edges or finer partitions of the color classes. The true graph $G_T$ contains the edge set $E_T$, the edge color classes $E_1^T, \cdots,E_k^T$ and the vertices color classes $V_1^T, \cdots,V_t^T$. The competitor graph $G_s$ contains the edge set $E_s$, the edge color classes $E_{1}^s, \cdots,E_{k}^s$ and the vertices color classes $V_{1}^s, \cdots,V_{t}^s$. If the following three conditions are satisfied and $G_s\neq G_T$, then the graph $G_s$ is called an overfitting graph denoted by $G_+$. Any other competitor graph is called an underfitting graph which is denoted by $G_-$.

\begin{enumerate}
\item $E_T \subseteq  E_s$.
\item For any edge class $E^s_i, i\in \{1,\cdots, k^s\}$ in $G^s$, there exists a edge class $E^T_j, j\in \{1,\cdots, k\}$ in $G^T$ such that
$$E^s_i\cap E_T \subseteq E^T_j.$$
\item For any vertex class $V^s_i, i\in \{1,\cdots, t^s\}$ in $G^s$, there exists a vertex class $V^T_j, j\in \{1,\cdots, t\}$ in $G^T$ such that
$$V^s_i \subseteq V^T_j.$$
\end{enumerate}
Denote $d_G$ as the number of parameters in the colored graphical model with underlying graph $G$, i.e., the total number of edge classes and vertex classes.  Then by the definition above, $d_{G_+} > d_{G_T}$.
The following theorem demonstrates that, as the sample size $n$ tends to infinity, the Bayes factor is model selection consistent for colored Gaussian graphical models.
\begin{thm}
\label{consistency}
Consider a colored graphical model where the number of vertices in the underlying graph is finite. Let $G_T$ be the true graph and let $G_s$ be any competing graph, then
$$\lim_{n\rightarrow \infty} p_T\Big( p(G_T|\textbf{X})>p(G_s|\textbf{X})\Big)=1.$$
\end{thm}

%The theorem shows the ratio of the posterior probabilities, \textcolor{blue}{i.e. the Bayes factor,} is model selection consistent. \textcolor{blue}{ Thus our model selection process, where each decision is made on the basis of the Bayes factor between the current graph and a competitor graph, is model selection consistent.} According to the Theorem \ref{consistency} above, asymptotically, the ratio of the posterior probabilities is maximized at the true graph $G.$ Our iterative search algorithm is a hill climbing algorithm on the posterior probability.\textcolor{blue}{ Did we prove this last statement? The theorem says that asymptotically we choose the true graph but how does it relate to our selection procedure? Our procedure is step by step comparing two graphs of which neither is the true graph. I am having trouble making the link between our procedure and the proof of Theorem 5.1.}

Theorem 5.1 states that,
under the true model $G_T$, the probability that the Bayes factor for the true model $G_T$ versus any other model $G$ be greater than one, tends to one as the sample size increases. At each step of our procedure, we move from one graph $G^{[t]}$ to another graph $G^*$ if and only if the Bayes factor for $G^*$ versus $G^{[t]}$ is greater than one. So, though we cannot guarantee that our procedure converges to the true graph $G_T$, we see that it moves towards the true graph.

If sample size goes to infinity and we use Bayes factor to compare all candidate models, we will be able to find the true model with probability tending to one. However due to the astronomical size of the model space, we cannot evaluate and compare all possible models. Thus our hill climbing algorithm is a practical strategy to search through the model space.

\section{Numerical experiments}
In subsection 6.1 below, we compare the performance of the proposed reversible jump algorithm and the double reversible jump algorithm to evaluate the Bayes factors. The values obtained are compared to the true values of the Bayes factor which, in the case of the small graphs that we consider, can be computed analytically and thus exactly numerically.
 In subsection 6.2, we analyze the validity of the proposed model selection approach based on the double reversible jump algorithm.

  For the prior $G$-Wishart distribution, the hyperparameters are $\delta = 3$ and $\textbf{D}=\textbf{I}$, the identity matrix. We use $\sigma = 0.5$ to generate the samples from the normal distributions and truncated normal distributions. We implement all algorithms in R.

\subsection{Estimation of Bayes factors}

\subsubsection{Graph with 3 vertices} 
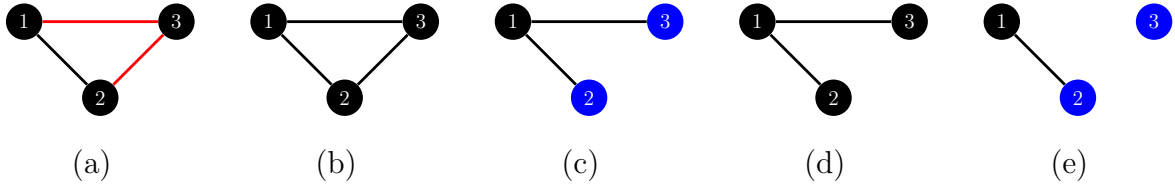
\begin{figure}
        \centering
        \begin{minipage}{.2\textwidth}
            \centering
            \vspace{9mm}
            \resizebox{0.8\textwidth}{!}{
\begin {tikzpicture}[circle]
\node[fill = black](C){\textcolor{white}{$2$}};
\node[fill = black] (A) [above left=of C] {\textcolor{white}{$1$}};
\node[fill = black] (B) [above right =of C] {\textcolor{white}{$3$}};
\draw[-][black,  ultra thick] (A) -- (C);
\draw[-][red,  ultra thick] (B) -- (C);
\draw[-][red,  ultra thick] (A) -- (B);
\end{tikzpicture}}

            (a)
 \label{fig:rect}
        \end{minipage}% \\
         \begin{minipage}{.2\textwidth}
            \centering
            \vspace{9mm}
            \resizebox{0.8\textwidth}{!}{
            \begin {tikzpicture}[circle]
\node[fill = black](C){\textcolor{white}{$2$}};
\node[fill = black] (A) [above left=of C] {\textcolor{white}{$1$}};
\node[fill = black] (B) [above right =of C] {\textcolor{white}{$3$}};
\draw[-][black,  ultra thick] (A) -- (C);
\draw[-][black,  ultra thick] (B) -- (C);
\draw[-][black,  ultra thick] (A) -- (B);
\end{tikzpicture}}

(b)
 \label{fig:rect}
        \end{minipage}% \\
         \begin{minipage}{.2\textwidth}
            \centering
            \vspace{9mm}
            \resizebox{0.8\textwidth}{!}{
            \begin {tikzpicture}[circle]
\node[fill = blue](C){\textcolor{white}{$2$}};
\node[fill = black] (A) [above left=of C] {\textcolor{white}{$1$}};
\node[fill = blue] (B) [above right =of C] {\textcolor{white}{$3$}};
\draw[-][black,  ultra thick] (A) -- (C);
\draw[-][black,  ultra thick] (A) -- (B);
\end{tikzpicture}}

            (c)
            \label{fig:square}
        \end{minipage}%
        \begin{minipage}{.2\textwidth}
            \centering
            \vspace{9mm}
            \resizebox{0.8\textwidth}{!}{
\begin {tikzpicture}[circle]
\node[fill = black](C){\textcolor{white}{$2$}};
\node[fill = black] (A) [above left=of C] {\textcolor{white}{$1$}};
\node[fill = black] (B) [above right =of C] {\textcolor{white}{$3$}};
\draw[-][black,  ultra thick] (A) -- (C);
\draw[-][black,  ultra thick] (A) -- (B);
\end{tikzpicture}}

            (d)
 \label{fig:rect}
        \end{minipage}% \\
         \begin{minipage}{.2\textwidth}
            \centering
            \vspace{9mm}
            \resizebox{0.8\textwidth}{!}{
            \begin {tikzpicture}[circle]
\node[fill = blue](C){\textcolor{white}{$2$}};
\node[fill = black] (A) [above left=of C] {\textcolor{white}{$1$}};
\node[fill = blue] (B) [above right =of C] {\textcolor{white}{$3$}};
\draw[-][black,  ultra thick] (A) -- (C);
\end{tikzpicture}}

            (e)
\label{fig:rect}
        \end{minipage}% \\
\caption{Colored graphs with 3 vertices.}
\label{fig:1}
\end{figure}

This simulation experiment is designed to assess the accuracy of the estimates of Bayes factors produced by our proposed samplers.  Since in the case of graphs with three vertices, we can compute the normalizing constant of the $G$-Wishart distribution. We empirically compare the estimates of Bayes factors using the reversible jump MCMC (RJ), the double reversible jump MCMC (DRJ), and the theoretical value of the ratio of normalizing constants (RN). For two neighboring colored graphs $G_{1}$ and $G_{2}$, the RN is defined by
$$RN = \frac{\frac{I_{G_{1}}(\delta+n, \textbf{D}+\textbf{S})}{I_{G_{1}}(\delta, \textbf{D})}}{\frac{I_{G_{2}}(\delta+n, \textbf{D}+\textbf{S})}{I_{G_{2}}(\delta, \textbf{D})}}.$$
We consider different colored graphs with three vertices in Figure \ref{fig:1} for which we can compute the corresponding normalizing constants exactly. The normalizing constants of colored $G$-Wishart distributions for the five colored graphs shown in Figure \ref{fig:1} can be computed using Theorems 3 and 5 in Massam et al. (2018), Theorems 6.1, and the formula in \citet{Ata05}. We generate one dataset comprising $n=100$ observations sampled from the multivariate normal $N(0, \textbf{K}^{-1})$. The two MCMC samplers are run for 10,000 iterations with a burn-in of 1000 iterations.  Computational results are summarized in Tables \ref{table:1}, \ref{table:2} and \ref{table:3}. Results indicate that the double reversible jump MCMC can produce almost the same accuracy as the reversible jump MCMC. Estimates from both algorithms perform very well when compared with the theoretical values of Bayes factors.

The following theorem gives the analytic expression of the normalizing constants for  the colored $G$-Wishart distributions with underlying graph as shown in Figure \ref{fig:1} (e).

\begin{thm}
For the colored graph $G$ of Figure \ref{fig:1} (e), the normalizing constant $I_{G}(\delta, \textbf{D})$ is
\begin{eqnarray*}
I_{G}(\delta, \textbf{D})
&=& \frac{2^{\frac{\delta+1}{2}}\Gamma(\frac{\delta}{2})\sqrt{\pi}d_{11}^{-\frac{\delta+1}{2}}\Gamma(\delta-\frac{1}{2})}{[\frac{1}{2}(d_{22}
+d_{33})-\frac{d^2_{12}}{2d_{11}}]^{\delta-\frac{1}{2}}},
\end{eqnarray*}
where $\textbf{D} = (d_{ij})_{1\leq i,j \leq p}$.
\end{thm}
The proof of Theorem 6.1 is given in the Appendix.

%We implemented the MCMC algorithms in R.

\begin{table*}
\centering
\caption{Estimation of Bayes factors for Figure \ref{fig:1} (c) vs. Figure \ref{fig:1} (e). Standard errors are indicated in parentheses.}
\begin{tabular}{ccccc}
\hline
$\textbf{K}$ &$\begin{pmatrix}
     1 & 0.4 &0.2\\
     0.4 & 1 & 0\\
     0.2 &0 & 1\\
\end{pmatrix}$ &$\begin{pmatrix}
     1 & 0.4 &0.22\\
     0.4 & 1 &0\\
     0.22&0 & 1\\
\end{pmatrix}$& $\begin{pmatrix}
     1 & 0.4 &0\\
     0.4 & 1 & 0\\
     0 &0 & 1\\
\end{pmatrix}$& $\begin{pmatrix}
     1 & 0.4 &0.1\\
     0.4 & 1 & 0\\
     0.1 &0 & 1\\
\end{pmatrix}$
\\
RJ & 2.167 (0.122) &2.916 (0.084) &0.237 (0.005) &1.445 (0.034)\\
DRJ&2.114 (0.055)&3.314 (0.167) &0.242 (0.009) & 1.481 (0.051)\\
RN&2.181&3.017 &0.234 &1.469\\
\hline
\end{tabular}
\label{table:1}
\end{table*}

\begin{table*}
\caption{Estimation of Bayes factors for Figure \ref{fig:1} (b) vs. Figure \ref{fig:1} (a). Standard errors are indicated in parentheses.}
\centering
\begin{tabular}{ccccc}
\hline
$\textbf{K}$ &$\begin{pmatrix}
     1 & 0.4 &0.2\\
     0.4 & 1 & 0\\
     0.2 &0 & 1\\
\end{pmatrix}$ &$\begin{pmatrix}
     1 & 0.4 &0.38\\
     0.4 & 1 &0\\
     0.38&0 & 1\\
\end{pmatrix}$& $\begin{pmatrix}
     1 & 0.4 &0\\
     0.4 & 1 & 0\\
     0 &0 & 1\\
\end{pmatrix}$& $\begin{pmatrix}
     1 & 0.4 &0.1\\
     0.4 & 1 & 0\\
     0.1 &0 & 1\\
\end{pmatrix}$
\\
RJ & 3.953 (0.327) &6.053 (0.648) &0.151 (0.004) &0.229 (0.007)\\
DRJ&4.217 (0.410)&5.782 (0.580) &0.149 (0.006) &0.227 (0.013)\\
RN& 4.349&5.511 &0.152 &0.229\\
\hline
\end{tabular}
\label{table:2}
\end{table*}

\begin{table*}
\caption{Estimation of Bayes factors for Figure \ref{fig:1} (d) vs. Figure \ref{fig:1} (c). Standard errors are indicated in parentheses.}
\centering
\begin{tabular}{ccccc}
\hline
$\textbf{K}$ &$\begin{pmatrix}
     1 & 0.4 &0.2\\
     0.4 & 1 & 0\\
     0.2 &0 & 1\\
\end{pmatrix}$ &$\begin{pmatrix}
     1 & 0.4 &0.2\\
     0.4 & 1.35 &0\\
     0.2&0 & 1\\
\end{pmatrix}$& $\begin{pmatrix}
     1 & 0.4 &0.2\\
     0.4 & 1.2 & 0\\
     0.2 &0 & 1\\
\end{pmatrix}$& $\begin{pmatrix}
     1 & 0.4 &0.2\\
     0.4 & 1 & 0\\
     0.2 &0 & 1.1\\
\end{pmatrix}$
\\
RJ &  0.112 (0.004)&1.793 (0.078) &0.568 (0.030) &0.413 (0.013)\\
DRJ&0.105 (0.006)&1.708 (0.116)&0.545 (0.025) &0.402 (0.019)\\
RN&0.109&1.988 &0.480 &0.388\\
\hline
\end{tabular}
\label{table:3}
\end{table*}

\subsubsection{Star graphs with 8 vertices} 
\begin{figure}
  \centering
  \begin{minipage}[b]{0.4\textwidth}
    \includegraphics[width=0.9\textwidth]{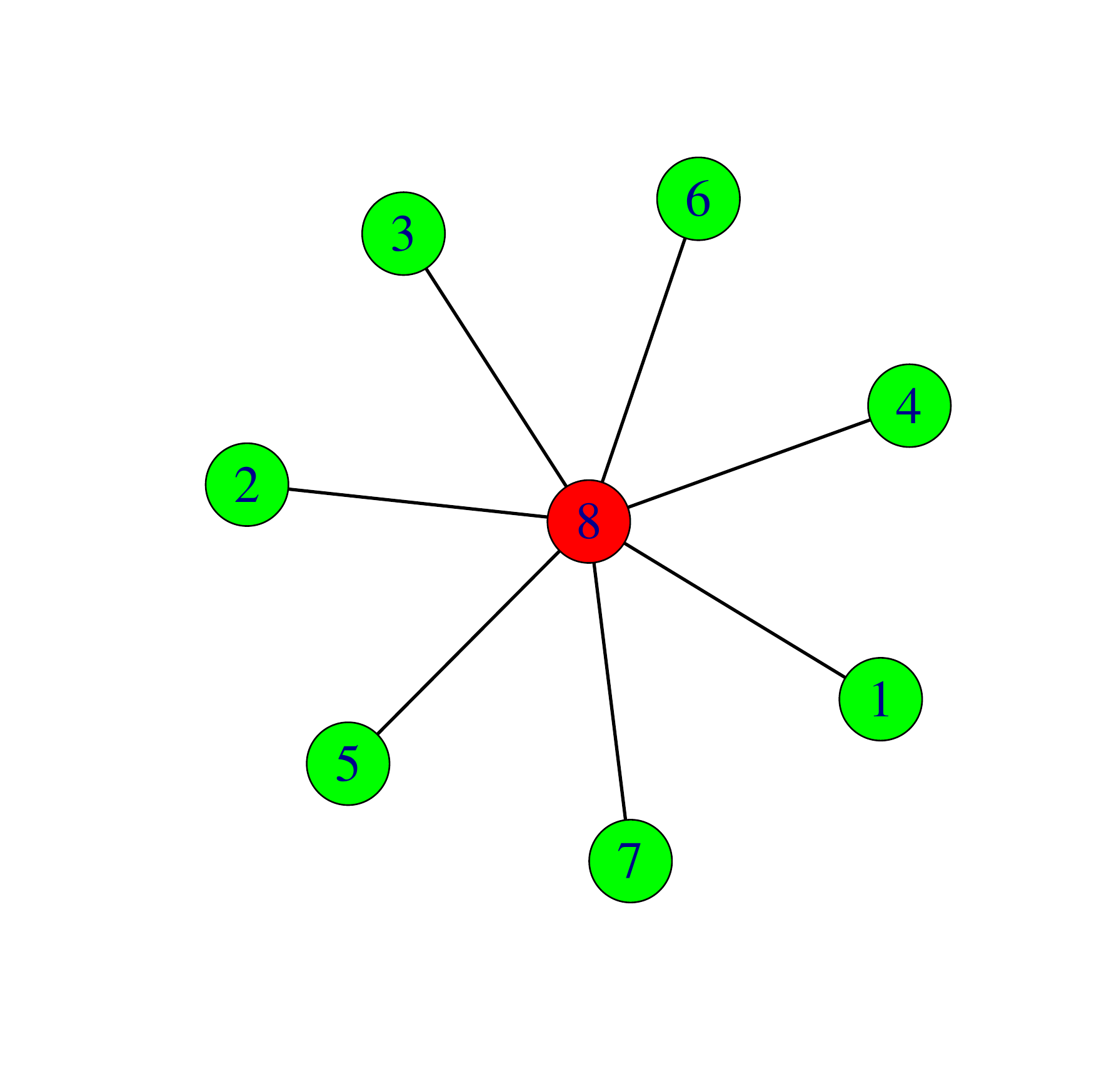}
\begin{center}
    (a)
\end{center}
  \end{minipage}
    \begin{minipage}[b]{0.4\textwidth}
    \includegraphics[width=0.9\textwidth]{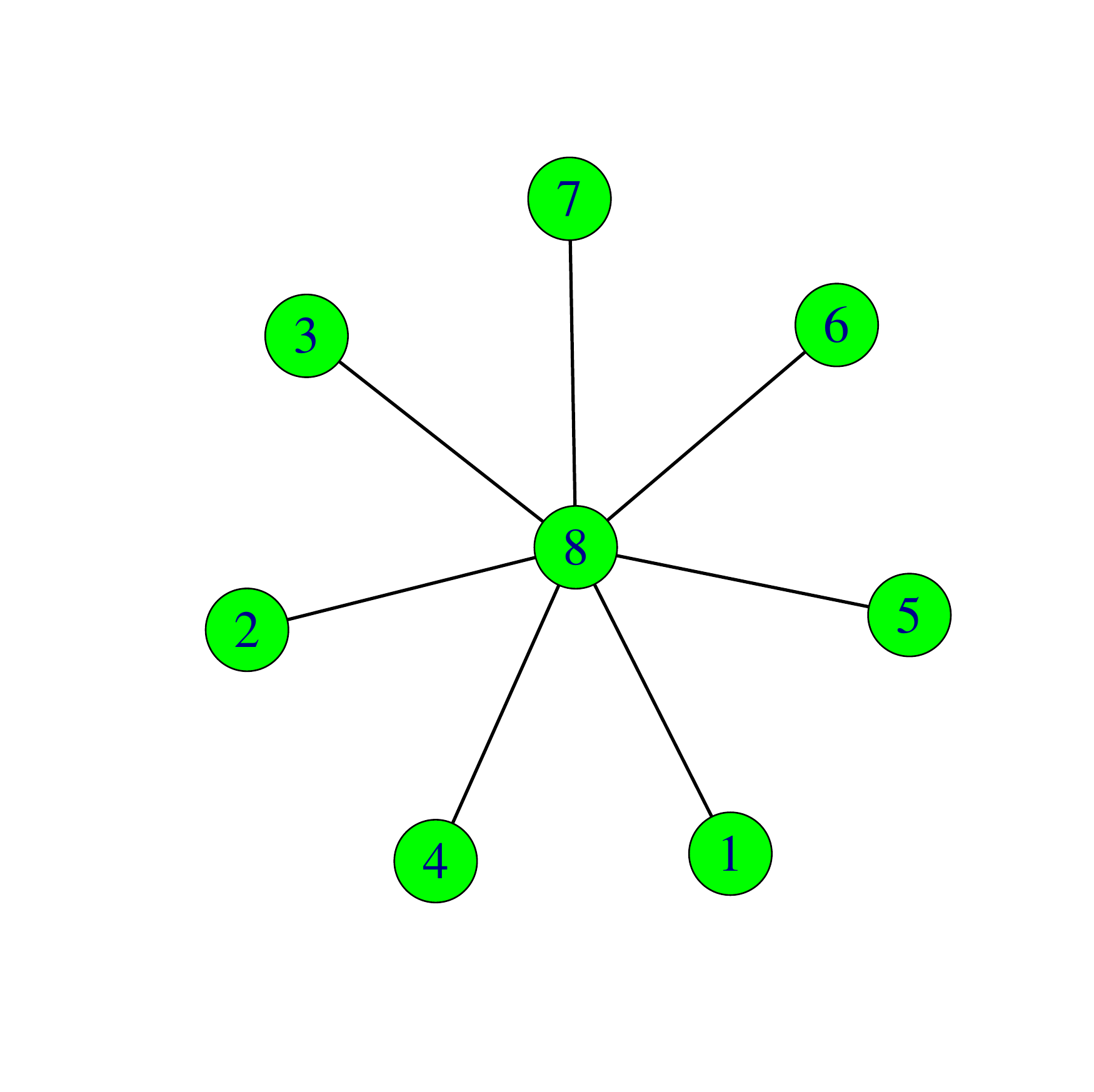}

\begin{center}
    (b)
\end{center}
  \end{minipage}
     \caption{Two star graphs with 8 vertices.}

     \label{fig:3}
\end{figure}

We consider two star graphs with $p=8$ in Figure \ref{fig:3}. The normalizing constants of colored $G$-Wishart distributions underlying the graphs in Figure \ref{fig:3} (a) and (b) are computed by Theorems 3 and 4 in \citet{Massam18}, respectively. Following \citet{Wang12}, we construct $\textbf{S}=n\textbf{K}^{-1}$ which corresponds to $n=100$ independent observations of $N(0,\textbf{K}^{-1})$. Furthermore, the precision matrix $\textbf{K}$ is given by $K_{ii}=1$ for $i=1,\cdots,p-1$ and $K_{pj}=K_{jp}=0.3$ for $j=1,\cdots,p-1$. The element $K_{pp}$ corresponding to the center of star graphs is shown in Table \ref{table:8}. To evaluate the performance of RJ and DRJ algorithms, we run these algorithms with 10000 iterations and 1000 as a burn-in. Table \ref{table:8} reports comparisons of our methods with the true values of the Bayes factors.

\begin{table*}
\caption{Estimation of Bayes factors for Figure \ref{fig:3} (a) vs. Figure \ref{fig:3} (b). Standard errors are indicated in parentheses.}
\centering
\begin{tabular}{ccccc}
\hline
$K_{pp}$ &1 &1.2& 1.3& 1.4
\\
RJ & 0.004 (0.001) &0.020 (0.002) &0.057 (0.005) &0.255 (0.026)\\
DRJ&0.008 (0.001)&0.020 (0.005) &0.051 (0.009) &0.199 (0.034)\\
RN&0.004&0.019 & 0.063 &0.230\\
\hline
\end{tabular}
\label{table:8}
\end{table*}

\subsection{Model selection}

We conduct model selection experiments using the algorithm proposed in Section 3. The double reversible jump MCMC algorithm is used to estimate the Bayes factors and the sampler is run for 5,000 iterations with a burn-in of 1000 iterations. The threshold for the linear regression step of our method is $\alpha=0,05$.

\subsubsection{Graphs with three vertices} We simulate 50 datasets each comprising $n=100$ observations sampled from multivariate normal $N(0, \textbf{K}^{-1})$. Tables \ref{table:4}, \ref{table:5} and \ref{table:6} show the average edge inclusion probabilities and percentages of the true graphical structure produced from the model selection algorithm. Our algorithm recovers the true graphical structure for different colored graphs very well: true colored models are selected with  probabilities above 0.82. The edges that should belong to the true graph receive inclusion probabilities above 0.90, while the edges that should be absent from the true graph receive inclusion probabilities below 0.08. The CPU time (in seconds) for one dataset is also given in Tables \ref{table:4}, \ref{table:5} and \ref{table:6}.

\begin{table*}
\caption{The percentages of the true model being selected over 50 simulated datasets, edge inclusion probabilities (IP) and timing (in seconds) for the algorithm of model selection.}
\centering
\begin{tabular}{ccccc}
\hline
$\textbf{K}$ &$\begin{pmatrix}
     1 & 0.4 &0.4\\
     0.4 & 1 & 0\\
     0.4 &0 & 1\\
\end{pmatrix}$ &$\begin{pmatrix}
     1 & 0.5 &0.5\\
     0.5 & 2 &0\\
     0.5&0 & 1\\
\end{pmatrix}$& $\begin{pmatrix}
     1 & 0.4 &0\\
     0.4 & 0.5 & 0.4\\
     0 &0.4 & 0.5\\
\end{pmatrix}$
\\
IP &$\begin{pmatrix}
     $*$ & 0.98 &0.98\\
     $*$ & $*$ & 0.02\\
     $*$ &$*$ & $*$\\
\end{pmatrix}$ &$\begin{pmatrix}
     $*$ & 1 &1\\
     $*$ & $*$ & 0.1\\
     $*$ &$*$ & $*$\\
\end{pmatrix}$& $\begin{pmatrix}
     $*$ & 1 &0\\
     $*$ & $*$ & 1\\
     $*$ &$*$ & $*$\\
\end{pmatrix}$
\\
Percentage&0.98&0.88&1\\
Timing&261.36&316.06 &277.71 \\
\hline
\end{tabular}
\\
\label{table:4}
\end{table*}

\begin{table*}
\caption{The percentages of the true model being selected over 50 simulated datasets, edge inclusion probabilities (IP) and timing (in seconds) for the algorithm of model selection.}
\centering
\begin{tabular}{ccccc}
\hline
$\textbf{K}$ &$\begin{pmatrix}
     1 & 0.4 &0.4\\
     0.4 & 1 & 0.4\\
     0.4 &0.4 & 1\\
\end{pmatrix}$ &$\begin{pmatrix}
     1.5 & 1 &1\\
     1 & 1 &0\\
     1&0 & 3\\
\end{pmatrix}$& $\begin{pmatrix}
     1 & 0.5 &1.5\\
     0.5 & 0.5 & 0\\
     1.5 &0 & 3\\
\end{pmatrix}$
\\
IP &$\begin{pmatrix}
     $*$ & 1 &0.94\\
     $*$ & $*$ & 0.98\\
     $*$ &$*$ & $*$\\
\end{pmatrix}$ &$\begin{pmatrix}
     $*$ & 1 &1\\
     $*$ & $*$ & 0\\
     $*$ &$*$ & $*$\\
\end{pmatrix}$& $\begin{pmatrix}
     $*$ & 0.06 &1\\
     $*$ & $*$ & 0.08\\
     $*$ &$*$ & $*$\\
\end{pmatrix}$
\\
Percentage&0.82&0.98&1\\
%Percentage(n=500)&&&0.86\\
Timing&346.64&308.56 &211.42 \\
\hline
\end{tabular}
\label{table:5}
\end{table*}

\begin{table*}
\caption{The percentages of the true model being selected over 50 simulated datasets, edge inclusion probabilities (IP) and timing (in seconds) for the algorithm of model selection.}
\centering
\begin{tabular}{ccccc}
\hline
$\textbf{K}$ &$\begin{pmatrix}
     0.5 & 0 &0\\
     0 & 2 & 0\\
     0 &0 & 0.5\\
\end{pmatrix}$ &$\begin{pmatrix}
     0.5 & 0 &0\\
     0 & 0.5 &0\\
     0&0 & 0.5\\
\end{pmatrix}$& $\begin{pmatrix}
     1.5 & 0.6 &0\\
     0.6 & 2 & 1.4\\
     0 &1.4 & 1.5\\
\end{pmatrix}$
\\
IP &$\begin{pmatrix}
     $*$ & 0.08 &0.06\\
     $*$ & $*$ & 0.06\\
     $*$ &$*$ & $*$\\
\end{pmatrix}$ &$\begin{pmatrix}
     $*$ & 0.06 &0.06\\
     $*$ & $*$ & 0.04\\
     $*$ &$*$ & $*$\\
\end{pmatrix}$& $\begin{pmatrix}
     $*$ & 0.9 &0.06\\
     $*$ & $*$ & 1\\
     $*$ &$*$ & $*$\\
\end{pmatrix}$
\\
Percentage&0.84&0.86&0.84\\
Timing&346.64&308.56 &211.42 \\
\hline
\end{tabular}
\label{table:6}
\end{table*}

\subsubsection{Star graphs with $p$=8 vertices and $p$=11 vertices} Let $\textbf{K}^{T}$ and $\hat{\textbf{K}}$ be the true precision matrix and the  precision matrix selected through the model selection approach, respectively.
%Suppose there are $s$ vertex color classes and $t$ edge color classes in the true graph, let $V_{1},\cdots,V_{s}$ be the sets of vertices in which the corresponding $K^{0}_{ii}=K^{0}_{jj}$ if $(i,i)\in V_{k}$, $(j,j)\in V_{k}$ for some $k=1,\cdots,s,$ and $i\neq j$. Similarly, let $E_{1}, \cdots, E_{t}$ be the sets of edges in which the corresponding $K^{0}_{ij}=K^{0}_{i'j'}$ if $(i,j) \in E_{k}$, $(i',j') \in E_{k}$ for some $k=1,\cdots,t,$ and $(i,j)\neq(i',j')$.
To assess the performance of the symmetric structure, we compute the measures $d_{0}$, $d_{V^{T}_{i}}$, $i=1,\cdots,t$, $d_{E^{T}_{j}}$, $j=1,\cdots,k$, and $Acc_{all}$  defined below for measuring supervised clustering and feature selection in \citet{Shen12}. Let $B$ be the set of missing edges in the true graph, i.e. such that $K^{T}_{ij}=0.$

 We define
$$d_{0} = \{\sum\limits_{(i,j)\in B}{\bf 1}_{\hat{K}_{ij}=0}+\sum\limits_{(i,j)\notin B}{\bf 1}_{\hat{K}_{ij}\neq 0}\}\Big/\frac{p(p-1)}{2}$$
concerning the zero constraints. For $i=1,\cdots,t$, let
$$d_{V^{T}_{i}} = \frac{\sum\limits_{j\neq j':(j,j)\in V^{T}_{i},(j',j')\in V^{T}_{i}}{\bf 1}_{\hat{K}_{jj}=\hat{K}_{j'j'}}+\sum\limits_{j\neq j':(j,j)\in V^{T}_{i},(j',j')\notin V^{T}_{i}}{\bf 1}_{\hat{K}_{jj}\neq \hat{K}_{j'j'}}}{|V^{T}_{i}|(p-1)}$$
which measures the performance in identifying the true vertex color classes. For $j=1,\cdots,k$, let
$$d_{E^{T}_{j}} = \frac{\sum\limits_{\substack{(k,l)\neq (k',l'):\\(k,l)\in E^{T}_{j},\\(k',l')\in E^{T}_{j}}}{\bf 1}_{\hat{K}_{kl}=\hat{K}_{k'l'}}+\sum\limits_{\substack{(k,l)\neq (k',l'):\\(k,l)\in E^{T}_{j},\\(k',l')\notin E^{T}_{j}}}{\bf 1}_{\hat{K}_{kl}\neq \hat{K}_{k'l'}}}{|E^{T}_{j}|[\frac{p(p-1)}{2}-1]}$$
which measures the performance in identifying the true edge color classes. We further define
$$Acc_{all} = \frac{d_{0}+\sum\limits^{s}_{i=1}d_{V^{T}_{i}}+\sum\limits^{t}_{j=1}d_{E^{T}_{j}}}{1+s+t}.$$ Note that $Acc_{all}$ lies between 0 and 1. The bigger $Acc_{all}$ is, the better performance is.

We simulate 20 datasets each comprising $n=1000$ observations sampled from multivariate normal $N(0, (\textbf{K}^{T})^{-1})$ where $K^T_{ii}=1$, $K^T_{ip}=K^T_{pi}=0.3$, for $i=1,2,\cdots,p$, and $K^T_{ij}=0$ for others.
%in which the precision matrix is
%
%\begin{center}
%$K=\begin{pmatrix}
%      1 & 0 &0&0&0&0&0&0.3\\
%      & 1 & 0&0&0&0&0&0.3\\
%      & & 1&0&0&0&0&0.3\\
%      & & &1 &0&0&0&0.3\\
%      & & &&1&0&0&0.3\\
%      &&&&&1&0&0.3\\
%     & & &&&&1&0.3\\
%     & & &&&&&1\\
%\end{pmatrix}.$
%\end{center}
%
%
%
%
%\[\left( \begin{array}{lllllllllll}
%       1 & 0 &0&0&0&0&0&0&0&0&0.3\\
%       0 & 1 &0&0&0&0&0&0&0&0&0.3\\
%       0 & 0 &1&0&0&0&0&0&0&0&0.3\\
%       0 & 0 &0&1&0&0&0&0&0&0&0.3\\
%       0 & 0 &0&0&1&0&0&0&0&0&0.3\\
%       0 & 0 &0&0&0&1&0&0&0&0&0.3\\
%       0 & 0 &0&0&0&0&1&0&0&0&0.3\\
%       0 & 0 &0&0&0&0&0&1&0&0&0.3\\
%       0 & 0 &0&0&0&0&0&0&1&0&0.3\\
%       0 & 0 &0&0&0&0&0&0&0&1&0.3\\
%       0 & 0 &0&0&0&0&0&0&0&0&1\\
%       \end{array} \right). \]
For $p=8$ and $p=11$, the average edge inclusion probabilities over 20 simulated datasets are

\begin{center}
$IP=\begin{pmatrix}
      $*$ & 0.05 &0.15&0.05&0.05&0.05&0.05&1\\
      &  $*$  & 0&0.05&0&0&0&1\\
      & & $*$&0.1&0&0&0&1\\
      & & &$*$ &0.05&0&0.05&0.95\\
      & & &&$*$&0&0.05&1\\
      &&&&&$*$&0.05&1\\
     & & &&&&$*$&1\\
     & & &&&&&$*$\\
\end{pmatrix}$
\end{center}
and
\[IP=\left( \begin{array}{lllllllllll}
       $*$ & 0 &0.1&0.05&0.2&0.1&0.05&0&0&0.05&1\\
         & $*$ &0.15&0&0.05&0.1&0.05&0&0&0.05&1\\
         &   &$*$&0.05&0&0.05&0.05&0&0.05&0&0.9\\
         &   &&$*$&0&0.1&0&0.05&0&0&0.9\\
         &   &&&$*$&0&0.05&0.1&0.1&0.05&0.85\\
         &   &&&&$*$&0&0.05&0.1&0&0.85\\
         &   &&&&&$*$&0&0.1&0&0.85\\
         &   &&&&&&$*$&0&0&0.9\\
         &   &&&&&&&$*$&0&0.75\\
         &   &&&&&&&&$*$&0.95\\
         &   &&&&&&&&&$*$\\
       \end{array} \right), \]
respectively. The values for measuring the symmetric structure are shown in Table \ref{table:measures}.

\begin{table*}
\caption{Measures for star graphs when $p=8$ and $p=11$. Standard errors are indicated in parentheses.}
\centering
\begin{tabular}{ccccc}
\hline
 &$\bar{d}_{0}$ &$\bar{d}_{V_{1}}$& $\bar{d}_{E_{1}}$& $\overline{Acc}_{all}$
\\
$p=8$ & 0.971(0.021) &0.879(0.164) &0.967(0.056) &0.940(0.066)\\
$p=11$&0.947(0.033)&0.383(0.222) &0.835(0.063) &0.722(0.103)\\
\hline
\end{tabular}
\label{table:measures}
\end{table*}

%For $p=8$, we obtain the averages $\bar{d}_{0}=0.971(0.021)$, $\bar{d}_{V_{1}}=0.879(0.164)$, $\bar{d}_{E_{1}}=0.967(0.056)$ and $\overline{Acc}_{all}=0.940(0.066)$. For $p=11$, we obtain the averages $\bar{d}_{0}=0.947(0.033)$, $\bar{d}_{V_{1}}=0.383(0.222)$, $\bar{d}_{E_{1}}=0.835(0.063)$ and $\overline{Acc}_{all}=0.722(0.103)$. Standard errors are indicated in parentheses.

\section{Application to flow cytometry data}

We analyze a flow cytometry dataset on signaling networks of human immune system cells. Flow cytometry can measure multiple molecules within each cell and it is possible to identify complex causal influence relationships involving multiple proteins \citep{Sachs05}. \citet{Sachs05} fitted a directed acyclic graph to depict the signalling pathway between the proteins. Using the fractional pseudo-likelihood method, \citet{Leppa17} analyzed interactions between the proteins based on the uncolored and undirected graphs. The dataset includes $p=11$ proteins and $n=7466$ observations.

We want to detect the underlying colored graph for the flow cytometry data using the proposed model selection algorithm in Section 3. Even with a moderate number of variables $p=11$, the model space for colored graphical models is astronomical in size. In order to speed up the model search approach, we add a likelihood ratio test \citep{Hoj08} before using Bayes factors estimation between two colored graphs. When comparing  a colored graph $G_{1}$ with a candidate graph $G_{2}$ obtained by adding one edge or deleting one vertex color class on $G_{1}$, we compute a likelihood ratio on one degree of freedom using the R function rcox() in the R package gRc \citep{Hoj07}. If the candidate graph $G_{2}$ is not selected using the likelihood ratio test, we will not consider $G_{2}$ in the model selection algorithm.

\begin{figure}
  \centering
    \includegraphics[width=0.5\textwidth]{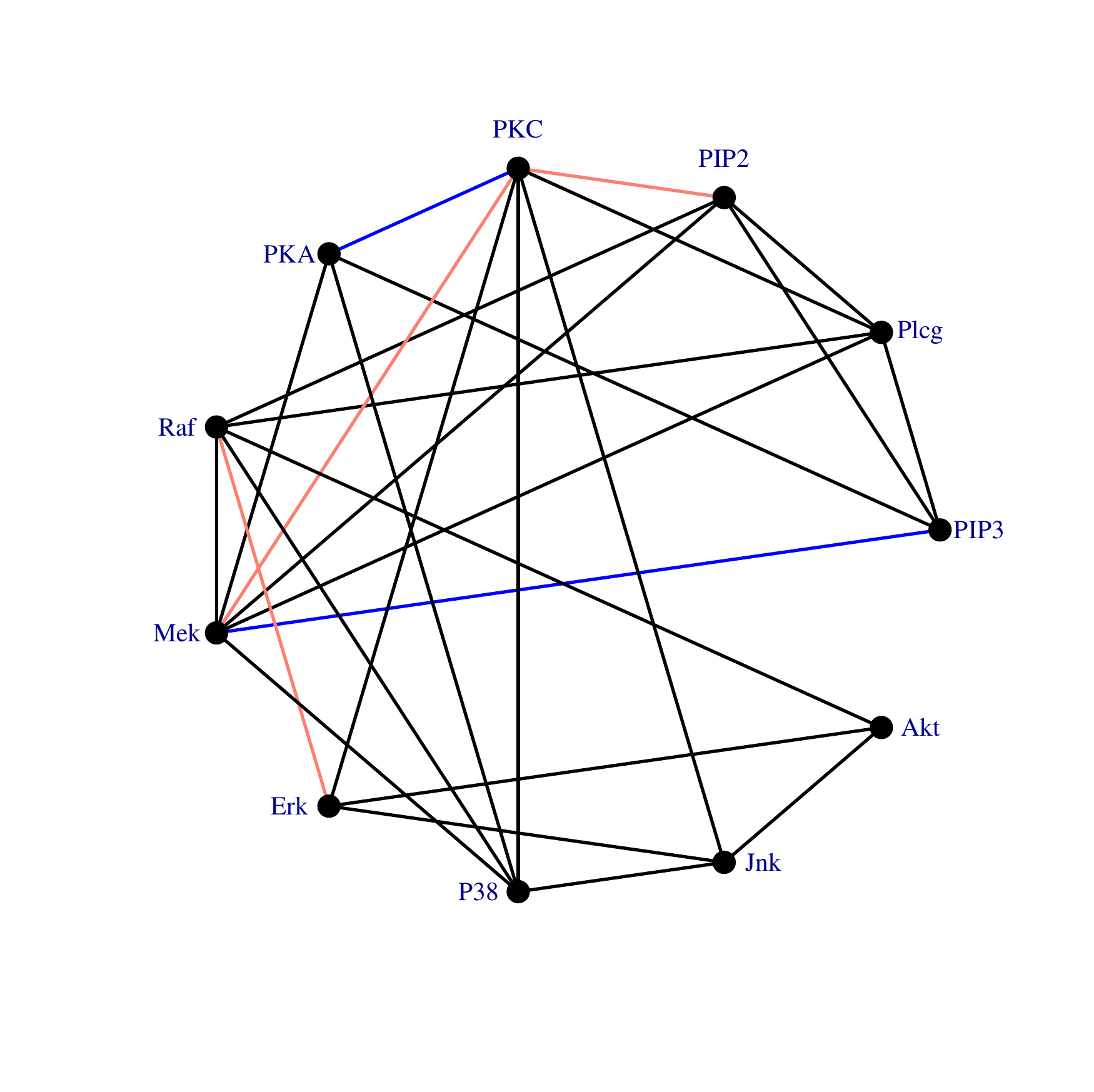}
    \caption{The best colored graph from flow cytometry data.}
    \label{fig:2}
\end{figure}

In the model selection procedure, we set $\delta=3$, $\sigma=0.5$ and $\textbf{D}$ be the identity matrix. Further, we run the double reversible jump algorithm for 10,000 iterations and discard the first 1,000 as burn-in. The best colored graph selected through our model selection algorithm is presented in Figure \ref{fig:2}. The selected graph contains 27 edges, 24 edge color classes and 11 vertex color classes. There are 18 edges in common with the edges in the  uncolored graph developed in \citet{Leppa17}, which has 31 edges in total.  Comparing with the results of \citet{Leppa17}, we detect many common edges and those warrant future biological validations. Our method has the advantage of detecting symmetries in the graphical model and in this particular case, we find that there are in fact only few symmetries in the graph.

\section{Concluding remarks}
We propose a Bayesian method to perform model selection in the class of colored Gaussian graphical models which allows us to capture the symmetric structure and dependency patterns of random variables.  Also, symmetry restrictions imposed on graphical Gaussian models reduce the number of parameters, which is useful  when the number of variables exceeds the number of observations.

We develop a trans-dimensional double reversible jump algorithm, which provides an accurate estimate of Bayes factors. Combining linear regression with the double reversible jump algorithm, we develop an efficient model selection approach for colored graphical models. Our  model selection approach avoids the calculation of normalizing constants of the colored $G$-Wishart distribution and saves  computing time when $p$ is large.
%
%In our model determination approach, the main computational part is sampling from the colored $G$-Wishart distribution of the precision matrix. The completion step for the Cholesky decomposition in the sampler methods is very time consuming as the dimension grows. In future work, we aim to improve the speed for high dimensional model search by parsing out a high-dimensional model to smaller models. Combing the selected model in smaller models to obtain the fitted model in a large graph will be the main idea for truly high dimensional problems.

%We provided an efficient model selection procedure for colored graphical models based on the colored $G$-Wishart distribution, although colored graphical models have more complex estimation properties. The proposed model selection methods, along with the algorithm for Bayes factors estimation, make the colored graphical models to become a promising tool in exploratory analysis of data.
%
%There are other techniques suitable for analyzing cell signalling data. It would be of interest to identify the symmetry structure of underling graphs which allows one to infer similarity in protein patterns. The proposed technique provide a useful visual tool to capture relationships between the measurements of proteins. On the other hand, the symmetry structure on the precision matrix displayed provides an alternative view of the cell signalling data.

\appendix

\section*{Appendix}\label{app}

This section contains the proofs of Theorems 5.1 and 6.1.

\subsection{Proof of Theorem 5.1}

Given any model $G$ with edge classes $E_{1}, \cdots,E_{k}$ and vertex classes $V_1, \cdots,V_t$, the posterior probability of $G$ given $\textbf{X}$ is:

\begin{eqnarray*}\label{marginal}
p(G|\textbf{X}) = \int_{\textbf{K}\in P_G} p(G,\textbf{K}|\textbf{X}) d \textbf{K}&\propto&\int_{\textbf{K}\in P_G} p(\textbf{X}|G,\textbf{K})p(\textbf{K}|G)p(G)d \textbf{K}.\\
%&=& \int_{K\in P_G} \frac{1}{(2\pi)^{p/2}}|K|^{n/2}\exp\big\{-\frac{1}{2} <K,S>
%\big\}1_{K \in P_{G}} \pi(K) p(G)d K.
\end{eqnarray*}

We approximate $p(G|\textbf{X})$ using the Laplace formula. For a given $\textbf{K}$ corresponding to the colored graph $G$, we rewrite $\textbf{K}$ as a $(k+t)$-dimensional vector $\boldsymbol{\theta}$ with $\boldsymbol{\theta}=(\theta_{V_1}, \cdots, \theta_{V_t}, \theta_{E_1}, \cdots, \theta_{E_k})$. Let $q(\boldsymbol{\theta},\textbf{X},G) = 1/n\times\log(p(\textbf{X}|G,\boldsymbol{\theta})p(\boldsymbol{\theta}|G))$ and $\tilde{\boldsymbol{\theta}}$ be the mode of $q(\boldsymbol{\theta},\textbf{X},G),$ where $p(G,\boldsymbol{\theta}|\textbf{X})$ attains the maximum value. Then we have
\begin{eqnarray*}\label{marginal}
p(G|\textbf{X}) &\propto& p(G)\int_{\boldsymbol{\theta}\in P_G} p(\textbf{X}|G,\boldsymbol{\theta})p(\boldsymbol{\theta}|G) d \boldsymbol{\theta} \\
 &=& p(G)\int_{\boldsymbol{\theta}\in P_G} \exp \{ n \frac{1}{n}\log \big(p(\textbf{X}|G,\boldsymbol{\theta})p(\boldsymbol{\theta}|G)\big)\}d \boldsymbol{\theta}\\
&=& p(G)\int_{\boldsymbol{\theta}\in P_G} \exp \{n q(\tilde{\boldsymbol{\theta}},\textbf{X},G)-\frac{n}{2}(\boldsymbol{\theta}-\tilde{\boldsymbol{\theta}})^\top Q(\mathbf{\boldsymbol{\theta}}^*,\textbf{X})(\boldsymbol{\theta}-\tilde{\boldsymbol{\theta}})\}d \boldsymbol{\theta}\\
&=& p(G)p(\textbf{X}|G,\tilde{\boldsymbol{\theta}})p(\tilde{\boldsymbol{\theta}}|G)(2\pi)^{\frac{d_G}{2}} n^{\frac{-d_G}{2}}|Q(\boldsymbol{\theta}^*,\textbf{X})|^{-1/2}\\
&=& \exp\Big\{\log p(G)+n q(\tilde{\boldsymbol{\theta}},\textbf{X},G)+\frac{1}{2}d_G\log 2\pi-\frac{1}{2}d_G\log n-\frac{1}{2}\log |Q(\boldsymbol{\theta}^*,\textbf{X})|\Big\},
%&=& \log p(G)+\log p(X|G,\tilde{\theta}) + \log p(\tilde{\theta}|G) + \frac{1}{2}d_G\log 2\pi- \frac{1}{2}d_G\log n-\frac{1}{2}\log |Q(\theta^*,X)|
\end{eqnarray*}
where $Q(\boldsymbol{\theta}^*,\textbf{X}) = -\partial^2 q(\boldsymbol{\theta},\textbf{X},G)/\partial \boldsymbol{\theta} \partial \boldsymbol{\theta}^\top|_{\boldsymbol{\theta}^*}$, and $\boldsymbol{\theta}^*$ is within a small neighborhood of $\tilde{\boldsymbol{\theta}}.$

Following the definition in \citet{Hoj08}, we define the adjacency matrix $\mathbf{T}^{u}$ for a edge or vertex color class $u$. For each vertex color class $u$, $u\in V_{i}, i= 1,\cdots,t$, we define an $|V|\times |V|$ diagonal matrix $\mathbf{T}^{u}$ with entries $T_{\alpha\alpha}=1$ if $\alpha \in u$ and 0 otherwise. Similarly, for each color class $u$, $u\in E_j, j=1,\cdots,k$, we let $\mathbf{T}^{u}$ be the $|V|\times |V|$ symmetric matrix with entries $T^{u}_{\alpha \beta} = 1$ if $(\alpha,\beta) \in u$ and 0 otherwise. We also have that
\begin{eqnarray*}
q(\boldsymbol{\theta},\textbf{X},G) &=& \frac{1}{n}\log(p(\textbf{X}|G,\textbf{K})p(\textbf{K}|G)) \\
&=& \frac{1}{n}\log\left\{ \frac{1}{I_{G}(\delta, \textbf{D})}|\textbf{K}|^{(n+\delta-2)/2}\exp\big\{-\frac{1}{2} <\textbf{K},\textbf{S}+\textbf{D}>
\big\}{\bf 1}_{P_{G}}(\textbf{K})\right\}\\
&=& \frac{1}{n}\left\{-\log I_{G}(\delta, \textbf{D}) + \frac{n+\delta-2}{2}\log |\textbf{K}|-\frac{1}{2}<\textbf{K},\textbf{S}+\textbf{D}>\right\}{\bf 1}_{P_{G}}(\textbf{K}) .
\end{eqnarray*}

By equation (6) in \citet{Hoj08},  we have that
$$
\frac{\partial^2 q(\boldsymbol{\theta},\textbf{X},G)}{\partial \theta^u \partial \theta^v}|_{\boldsymbol{\theta}^{*}} = -\frac{1}{n}\frac{n+\delta-2}{2}tr(\mathbf{T}^u\mathbf{\Sigma}^* \mathbf{T}^v\mathbf{\Sigma}^*)
$$
for two color classes $u$ and $v$, where $tr(\cdot)$ denotes the trace of a matrix, $\mathbf{\Sigma}^{*}=(\textbf{K}^*)^{-1}$ which corresponds to $\boldsymbol{\theta}^*$. Thus, the term $\log |Q(\boldsymbol{\theta}^*,\textbf{X})|$ is  of order $O_p(1).$

%(Note: to see the definition of $T^i$ and $T^j$ in Hojasgard (2008)
%\begin{eqnarray*}
%q(K,X,G) &=& \frac{1}{n}\log(p(X|G,K)p(K|G)) \\
%&=& \frac{1}{n}\log \frac{1}{I_{G}(\delta, D)}|K|^{(n+\delta-2)/2}\exp\big\{-\frac{1}{2} <K,S+D>
%\big\}1_{K \in P_{G}}. \\
%&=& \frac{1}{n}[-\log I_{G}(\delta, D) + \frac{n+\delta-2}{2}\log |K|1_{K \in P_{G}}-\frac{1}{2}<K,S+D>1_{K \in P_{G}}]
%\end{eqnarray*}
%Therefore, $$\frac{\partial q(K,X,G)}{\partial \theta^i} = \frac{1}{n}\frac{n+\delta-2}{2}tr(T^i\Sigma) - \frac{1}{2}(S+D) $$
%and $$\frac{\partial^2 q(K,X,G)}{\partial \theta^i \partial \theta^j} = -\frac{1}{n}\frac{n+\delta-2}{2}tr(T^i\Sigma T^j\Sigma)$$
%).{\bf this part should be presented in the proof, not in parenthesis.}

Let $\hat{\boldsymbol{\theta}}$ be the MLE. Now, we take the Taylor expansion of $\log p(\textbf{X}|G,\tilde{\boldsymbol{\theta}})$ around $\hat{\boldsymbol{\theta}},$
\begin{eqnarray*}
\log p(\textbf{X}|G,\tilde{\boldsymbol{\theta}}) = \log p(\textbf{X}|G,\hat{\boldsymbol{\theta}}) + \frac{1}{2}(\tilde{\boldsymbol{\theta}}-\hat{\boldsymbol{\theta}})^\top H(\boldsymbol{\theta}^{**},\textbf{X})(\tilde{\boldsymbol{\theta}}-\hat{\boldsymbol{\theta}}),
\end{eqnarray*}
where $\boldsymbol{\theta}^{**}$ is vector in a small neighborhood of $\hat{\boldsymbol{\theta}}$ and $H(\boldsymbol{\theta}^{**},\textbf{X})=-\partial^2 p(\textbf{X}|G,\boldsymbol{\theta})/\partial \boldsymbol{\theta}\partial \boldsymbol{\theta}^\top|_{\boldsymbol{\theta}^{**}}$.
Using similar arguments as for $Q(\boldsymbol{\theta}^*,\textbf{X})$, we have that $H(\boldsymbol{\theta}^{**},\textbf{X})$ is of order $O_p(n).$  According to
Theorem 4.3 in \citet{Ghosh06}, we have that
$\sqrt{n}(\tilde{\boldsymbol{\theta}}-\hat{\boldsymbol{\theta}})\to 0$ with a probability one. It implies $\tilde{\boldsymbol{\theta}}-\hat{\boldsymbol{\theta}} = O_p(1/\sqrt{n})$
and $(\tilde{\boldsymbol{\theta}}-\hat{\boldsymbol{\theta}})^\top H(\boldsymbol{\theta}^{**},\textbf{X})(\tilde{\boldsymbol{\theta}}-\hat{\boldsymbol{\theta}})=O_p(1)$. Thus, $\log p(\textbf{X}|G,\tilde{\boldsymbol{\theta}}) = \log p(\textbf{X}|G,\hat{\boldsymbol{\theta}})+O_p(1)$.

Let $\boldsymbol{\theta}_{T}$ be the parameter in the true model under which the data is generated. Let $\hat{\boldsymbol{\theta}}_T$, $\hat{\boldsymbol{\theta}}_s^+$ and $\hat{\boldsymbol{\theta}}_s^-$ be the MLEs corresponding to the true graph $G_T$, $G_{+}$ and $G_{-}$, respectively.
Consider an under-fitting model with the underlying graph $G_{-}$, we define a pseudo true value $\boldsymbol{\theta}^*_{-}$ with
$$\boldsymbol{\theta}^*_{-}=argmin_{\boldsymbol{\theta}_{-}}E_{\boldsymbol{\theta}^T}\{\log \frac{p(\textbf{X}|G_{T},\boldsymbol{\theta}_T)}{p(\textbf{X}|G_{-},\boldsymbol{\theta}_{-})}\}$$
under the misspecified model \citep{White82}.
By the weak Law of large numbers,
$$l(\hat{\boldsymbol{\theta}}_{-})/n \xrightarrow{\text{p}} E_{\boldsymbol{\theta}_T}\{\log p(\textbf{X}|G_{-},\boldsymbol{\theta}^*_{-})\},$$
$$l(\hat{\boldsymbol{\theta}}_{T})/n \xrightarrow{\text{p}} E_{\boldsymbol{\theta}_T}\{\log p(\textbf{X}|G_{T},\boldsymbol{\theta}_{T})\}.$$
This entails $l(\hat{\boldsymbol{\theta}}_{-})-l(\hat{\boldsymbol{\theta}}_{T})=  E_{\boldsymbol{\theta}_T}\{\log p(\textbf{X}|G_{-},\boldsymbol{\theta}^*_{-})\}- E_{\boldsymbol{\theta}_T}\{\log p(\textbf{X}|G_{T},\boldsymbol{\theta}_{T})\}+o_p(n).$
Furthermore, $E_{\boldsymbol{\theta}_T}\{\log p(\textbf{X}|G_{-},\boldsymbol{\theta}^*_{-})\}<E_{\boldsymbol{\theta}_T}\{\log p(\textbf{X}|G_{T},\boldsymbol{\theta}_{T})\}$ due to the Kullback-Leibler inequality. Combining the results above, we have
\begin{eqnarray*}
&&\log \frac{p(G_T\mid \mathbf{X})}{p(G_-\mid \mathbf{X})}\\
&=&\log\int_{\boldsymbol{\theta}\in P_{G_T}} p(\textbf{X}|G_T,\boldsymbol{\theta})p(\boldsymbol{\theta}|G_T) d \boldsymbol{\theta} - \log\int_{\boldsymbol{\theta}\in P_{G_{-}}} p(\textbf{X}|G_{-},\boldsymbol{\theta})p(\boldsymbol{\theta}|G_{-}) d \boldsymbol{\theta}\;\;\\
&=& \log p(\textbf{X}|G_T,\tilde{\boldsymbol{\theta}}_T) - \log p(\textbf{X}|G_{-},\tilde{\boldsymbol{\theta}}_{-}) + \log p(\tilde{\boldsymbol{\theta}}_T\mid G_T)-\log p(\tilde{\boldsymbol{\theta}}_-\mid G_-)\\
&& - \frac{1}{2}d_{G_T}\log n + \frac{1}{2}d_{G_{-}}\log n+O_p(1)\\
&=& [l(\hat{\boldsymbol{\theta}}_{T})-l(\hat{\boldsymbol{\theta}}_{-})] + \frac{1}{2}(d_{G_{-}}-d_{G_T})\log n + O_p(1)\\
&=& n\big(E_{\boldsymbol{\theta}_T}\{\log p(\textbf{X}|G_{T},\boldsymbol{\theta}_{T})\}-E_{\boldsymbol{\theta}_T}\{\log p(\textbf{X}|G_{-},\boldsymbol{\theta}^*_{-})\}\big)+ \frac{1}{2}(d_{G_{-}}-d_{G_T})\log n + o_p(n).
\end{eqnarray*}
The second last equality holds because $p(\boldsymbol{\theta}|G)$ is the colored $G$-wishart prior defined in \eqref{prior}, and  $\log p(\tilde{\boldsymbol{\theta}}|G)$ is a $O_p(1)$ term with respect to the sample size $n.$ In the last  equality, the first term is the dominating term. Due to the KL inequality above and the fact that $d_{G_-}<d_{G_T}$,  the last line is  asymptotically positive. Thus, we have $p(G_T|\textbf{X})  >  p(G_{-}|\textbf{X})$ with probability tending to one when $n$ goes to infinity.

Consider an over-fitting model with the underlying graph $G_{+}$, we have that
\begin{eqnarray*}
&&\log\int_{\boldsymbol{\theta}\in P_{G_T}} p(\textbf{X}|G_T,\boldsymbol{\theta})p(\boldsymbol{\theta}|G_T) d \boldsymbol{\theta} - \log\int_{\boldsymbol{\theta}\in P_{G_{+}}} p(\textbf{X}|G_{+},\boldsymbol{\theta})p(\boldsymbol{\theta}|G_{+}) d \boldsymbol{\theta}\;\;\\\\
&=& \log p(\textbf{X}|G_T,\tilde{\boldsymbol{\theta}}_{T}) - \log p(\textbf{X}|G_{+},\tilde{\boldsymbol{\theta}}_{+}) - \frac{1}{2}d_{G_T}\log n + \frac{1}{2}d_{G_{-}}\log n+O_p(1)\\
&=& [l(\hat{\boldsymbol{\theta}}_{T})-l(\hat{\boldsymbol{\theta}}_{+})]+ \frac{1}{2}(d_{G_{+}}-d_{G_T})\log n + O_p(1).
\end{eqnarray*}
According to the standard asymptotic theory for the loglikelihood ratio statistics, we have that $-2\{l(\hat{\boldsymbol{\theta}}_{T})-l(\hat{\boldsymbol{\theta}}_{+})\}\sim \chi^2_{d_{G_T}-d_{G_{+}}}=O_p(1)$. Therefore, the dominating term is the second term. As $d_{G_{+}} > d_{G_T}$, then $p(G_T|\textbf{X})  >  p(G_{+}|\textbf{X}) $ with probability tending to one when $n$ goes to infinity.

\subsection{Proof of Theorem 6.1}

We write the Cholesky decomposition of $\textbf{K}$ under the form $\textbf{K} = \mathbf{A}\mathbf{A}^\top$ with

\[ A_{ij} = \left\{
   \begin{array}{l l}
     \sqrt{a_{ii}} & \quad \text{if $i=j$,}\\
     -a_{ij} & \quad \text{if $i < j$.}
   \end{array} \right.\]
Thus, the entries of $\mathbf{A}\mathbf{A}^\top$ are given by
\[ (AA^\top)_{ij} = \left\{
   \begin{array}{l l}
     a_{ii}+\sum\limits_{l>i}a^2_{il} & \quad \text{if $i=j$,}\\
     -a_{ij}\sqrt{a_{jj}}+\sum\limits_{l> \max\{i,j\}}a_{il}a_{jl} & \quad \text{if $i < j$.}
   \end{array} \right.\]

For the colored graph shown in Figure \ref{fig:1} (e),  we denote the corresponding precision matrix by
\[\textbf{K} = \left( \begin{array}{ccc}
K_{11} & K_{12} & 0 \\
K_{12} & K_{22} & 0 \\
0 & 0& K_{22} \end{array} \right).\]
Since $K_{13} = 0$ and $a_{33}>0$, we have $a_{13}=0$. In addition, the conditions $K_{23}=0$ and $a_{33}>0$ imply $a_{23} = 0$. It follows that
\[\textbf{K} = \left( \begin{array}{ccc}
a_{11}+a^2_{12} & -a_{12}\sqrt{a_{22}} & 0 \\
-a_{12}\sqrt{a_{22}} & a_{22} & 0 \\
0 & 0& a_{33} \end{array} \right).\]
It can be seen from $K_{22}=K_{33}$ that $a_{22}=a_{33}$. Now we can write $\textbf{K}$ as
\[\textbf{K} = \left( \begin{array}{ccc}
a_{11}+a^2_{12} & -a_{12}\sqrt{a_{22}} & 0 \\
-a_{12}\sqrt{a_{22}} & a_{22} & 0 \\
0 & 0& a_{33} \end{array} \right)\]
and $|\textbf{K}|=a_{11}a^2_{22}$. On the other hand, the Jacobian of the transformation from $\textbf{K}$ to $\mathbf{A}$ is
$$\mathbf{J}=\bordermatrix{~& k_{11} & k_{12} & k_{22}  \cr
                  a_{11} & 1 & 0& 0\cr
                  a_{12} & *&-\sqrt{a_{22}} & 0\cr
                  a_{22} & *&* &1 \cr}.
$$
We obtain $|\mathbf{J}|=a^{1/2}_{22}$ and $$<\textbf{K},\textbf{D}> = d_{11}(a_{11}+a^2_{12})+d_{22}a_{22}+d_{33}a_{22}+2d_{12}(-a_{12}\sqrt{a_{22}}).$$
Therefore the normalizing constant $I_{G}(\delta,\textbf{D})$ can be written as
\begin{eqnarray*}
\label{eq:n1}
\int_\mathbf{A} a_{11}^{\frac{\delta-2}{2}}a_{22}^{\delta-3/2}\exp \big \{-\frac{1}{2}d_{11}a_{11}-\frac{1}{2}d_{11}a^2_{12}-\frac{1}{2}(d_{22}+d_{33})a_{22}+d_{12}a_{12}\sqrt{a_{22}}\big\}d\mathbf{A},
\end{eqnarray*}
where $a_{ii}>0$, $a_{ij} \in R$, $i<j$, and $d\mathbf{A}$ denotes the product of all differentials. In the above integral, we have that the integral with respect to $a_{11}$ is a gamma integral  with
\begin{eqnarray*}
\int^{\infty}_0 a^{\frac{\delta-2}{2}}_{11}\exp(-\frac{1}{2}d_{11}a_{11})da_{11}=2^{\frac{\delta}{2}}\Gamma(\frac{\delta}{2})d^{-\frac{\delta}{2}}_{11}.
\end{eqnarray*}
The integral with respect to $a_{12}$ is a Gaussian integral with
\begin{eqnarray*}
\int^{\infty}_{-\infty} \exp(-\frac{1}{2}d_{11}a^2_{12}+d_{12}\sqrt{a_{22}}a_{12})da_{12}=\frac{\sqrt{2\pi}}{\sqrt{d_{11}}}\exp(\frac{d^2_{12}a_{22}}{2d_{11}}).
\end{eqnarray*}
After some simplifications, we have
\begin{eqnarray*}
I_{G}(\delta, \textbf{D})&=& 2^{\frac{\delta}{2}}\Gamma(\frac{\delta}{2})d^{-\frac{\delta}{2}}_{11}\frac{\sqrt{2\pi}}{\sqrt{d_{11}}}\int_{0}^{\infty} a_{22}^{\delta-\frac{3}{2}} \exp
\big \{-\frac{1}{2}(d_{22}
+d_{33})a_{22}+\frac{d^2_{12}}{2d_{11}}a_{22}\big \} d a_{22}\\
&=& \frac{2^{\frac{\delta+1}{2}}\Gamma(\frac{\delta}{2})\sqrt{\pi}d_{11}^{-\frac{\delta+1}{2}}\Gamma(\delta-\frac{1}{2})}{[\frac{1}{2}(d_{22}
+d_{33})-\frac{d^2_{12}}{2d_{11}}]^{\delta-\frac{1}{2}}}.
\end{eqnarray*}

\subsection*{Acknowledgements}
The authors would like to acknowledge Sagnik Datta for his assistance in programming.

%\bigskip
%\begin{center}
%{\large\bf SUPPLEMENTARY MATERIAL}
%\end{center}

%\begin{description}
%
%\item[Title:] Brief description. (file type)
%
%\item[R-package for  MYNEW routine:] R-package containing code to perform the diagnostic methods described in the article. The package also contains all datasets used as examples in the article. (GNU zipped tar file)
%
%\item[HIV data set:] Data set used in the illustration of MYNEW method in Section~ 3.2. (.txt file)
%
%\end{description}

%\section{BibTeX}
%
%We hope you've chosen to use BibTeX!\ If you have, please feel free to use the package natbib with any bibliography style you're comfortable with. The .bst file agsm has been included here for your convenience.

\bibliographystyle{apalike}

\bibliography{BayesianModelSelectionRef}
\end{document}